\documentclass[aps,pra,showpacs,amssymb,superscriptaddress,nofootinbib,twocolumn]{revtex4-1}

\usepackage{overpic,color}

\usepackage{amsmath,amsfonts,amsthm,amssymb}
\usepackage{setspace}
\usepackage{amsmath}
\usepackage{appendix}
\usepackage[utf8]{inputenc}
\usepackage{bbm}
\usepackage{bm}
\usepackage{amsbsy}
\usepackage{dsfont}
\usepackage{graphicx}
\usepackage{epsfig}
\usepackage{epstopdf}
\usepackage{dsfont}
\usepackage{color}
\usepackage[bookmarks=true, colorlinks]{hyperref}

\makeatletter
\newcommand\org@hypertarget{}
\let\org@hypertarget\hypertarget
\renewcommand\hypertarget[2]{%
  \Hy@raisedlink{\org@hypertarget{#1}{}}#2%
  }
\makeatother

\usepackage[figure,table]{hypcap}
\usepackage{enumerate}
\usepackage[resetlabels]{multibib}

\hypersetup{
	bookmarksnumbered,
	pdfstartview={FitH},
	citecolor={darkgreen},
	linkcolor={darkred},
	urlcolor={darkblue},
	pdfpagemode={UseOutlines}}
\definecolor{darkgreen}{RGB}{50,190,50}
\definecolor{darkblue}{RGB}{0,0,190}
\definecolor{darkred}{RGB}{238,0,0}
\usepackage{soul}

\newcommand{\Ho}{$\mathrm{H_1}$}
\newcommand{\Ht}{$\mathrm{H_2}$}
\newcommand{\Vo}{$\mathrm{V_1}$}
\newcommand{\Vt}{$\mathrm{V_2}$}
\newcommand{\be}{\begin{equation}}
\newcommand{\ee}{\end{equation}}
\newcommand{\mean}[1]{\langle {#1} \rangle}
\newcommand{\ket}[1]{\left| {#1} \right\rangle}
\newcommand{\bra}[1]{\left\langle {#1} \right|}
\newcommand{\braket}[2]{\langle #1 | #2 \rangle }
\newcommand{\ketbra}[2]{\left| {#1} \right\rangle \!\! \left\langle {#2} \right|}
\newcommand{\prjct}[1]{\mathinner{|{#1}\rangle}\!\!\mathinner{\langle{#1}|}}
\newcommand{\ii}{\mathrm{i}}
\newcommand{\tr}[1]{\mbox{$\mathrm{Tr}\left(#1\right)$}}

\newcommand{\rfig}[1]{fig. \ref{#1}}

\begin{document}

\title{Indistinguishable photons from a trapped-ion quantum network node}

\vspace{20mm}

\author{M. Meraner$^\dagger$}
\affiliation{Institut f\"ur Quantenoptik und Quanteninformation,\\
	\"Osterreichische Akademie der Wissenschaften, Technikerstr. 21A, 6020 Innsbruck,
	Austria}
\affiliation{
	Institut f\"ur Experimentalphysik, Universit\"at Innsbruck,
	Technikerstr. 25, 6020 Innsbruck, Austria}
	
\author{A. Mazloom$^\dagger$}
\affiliation{Departement Physik, Universit\"at Basel, Klingelbergstrasse 82, CH-4056 Basel, Switzerland }

\author{V. Krutyanskiy$^\dagger$}
\affiliation{Institut f\"ur Quantenoptik und Quanteninformation,\\
	\"Osterreichische Akademie der Wissenschaften, Technikerstr. 21A, 6020 Innsbruck,
	Austria}

\author{\\V. Krcmarsky}
\affiliation{Institut f\"ur Quantenoptik und Quanteninformation,\\
	\"Osterreichische Akademie der Wissenschaften, Technikerstr. 21A, 6020 Innsbruck,
	Austria}
\affiliation{
	Institut f\"ur Experimentalphysik, Universit\"at Innsbruck,
	Technikerstr. 25, 6020 Innsbruck, Austria}

\author{J. Schupp}
\affiliation{Institut f\"ur Quantenoptik und Quanteninformation,\\
	\"Osterreichische Akademie der Wissenschaften, Technikerstr. 21A, 6020 Innsbruck,
	Austria}
\affiliation{
	Institut f\"ur Experimentalphysik, Universit\"at Innsbruck,
	Technikerstr. 25, 6020 Innsbruck, Austria}

\author{D. Fioretto}
\affiliation{
	Institut f\"ur Experimentalphysik, Universit\"at Innsbruck,
	Technikerstr. 25, 6020 Innsbruck, Austria}

\author{P. Sekatski}
\affiliation{Departement Physik, Universit\"at Basel, Klingelbergstrasse 82, CH-4056 Basel, Switzerland }

\author{T. E. Northup}
\affiliation{
	Institut f\"ur Experimentalphysik, Universit\"at Innsbruck,
	Technikerstr. 25, 6020 Innsbruck, Austria}

\author{N. Sangouard}
\affiliation{Institut de physique th\'{e}orique, Universit\'{e} Paris Saclay, CEA, CNRS, F-91191 Gif-sur-Yvette, France}
\affiliation{Departement Physik, Universit\"at Basel, Klingelbergstrasse 82, CH-4056 Basel, Switzerland }

\author{B. P. Lanyon}
\email{ben.lanyon@uibk.ac.at, $^\dagger$ These authors contributed equally}
\affiliation{Institut f\"ur Quantenoptik und Quanteninformation,\\
	\"Osterreichische Akademie der Wissenschaften, Technikerstr. 21A, 6020 Innsbruck,
	Austria}
	\affiliation{
	Institut f\"ur Experimentalphysik, Universit\"at Innsbruck,
	Technikerstr. 25, 6020 Innsbruck, Austria}

\begin{abstract}

Trapped atomic ions embedded in optical cavities are a promising platform to enable long-distance quantum networks and their most far-reaching applications. 
Here we achieve and analyze photon indistinguishability in a telecom-converted ion-cavity system. 
First, two-photon interference of cavity photons at their ion-resonant wavelength is observed and found to reach the limits set by spontaneous emission. 
Second, this limit is shown to be preserved after a two-step frequency conversion replicating a distributed scenario, in which the cavity photons are converted to the telecom C band and then back to the original wavelength. 
The achieved interference visibility and photon efficiency would allow for the distribution and practical verification of entanglement between ion-qubit registers separated by several tens of kilometers. 

\end{abstract}
\maketitle

Envisioned quantum networks, consisting of remote quantum matter linked up with light \cite{Kimble2008, Wehnereaam9288}, offer a fundamentally new communication paradigm \cite{Duan2001} as well as a practical path to large-scale quantum computation and simulation \cite{PhysRevA.89.022317} and to precision measurements in new regimes \cite{Komar2014, PhysRevLett.123.070504, sekatski2019optimal}. 
Trapped atomic ions are expected to enable the most promising applications of large-scale quantum networks \cite{Duan2010, Sangouard2009, RevModPhys.87.1379} given their demonstrated capabilities for quantum logic \cite{reviewionsqc}, multi-qubit registers \cite{Friis2018}, and optical clocks \cite{PhysRevLett.123.033201}.
Ion qubits have been entangled with propagating photons  \cite{Blinov2004} and those photons have been used to entangle ions in traps a few meters apart \cite{Moehring2007, Hucul:2015wo, balance}. 
Integrating ion traps with optical cavities offers the possibility of a near-deterministic and coherent light-matter interface for quantum networking \cite{RevModPhys.87.1379, Sangouard2009}, and both ion-photon entanglement \cite{Stute2012} and state transfer \cite{Stute:2013we} have been achieved in this setting. 

Recently, photons from trapped ions have been converted to the optimal telecom wavelengths for long-distance quantum networking via single-photon frequency conversion \cite{Bock2018, Walker2018, 50km}. In  \cite{50km} it was shown that the combination of an ion trap with an integrated optical cavity and telecom conversion could enable entanglement distribution between trapped ions spaced by tens of kilometres at practical rates for verification. 
However, it has not previously been verified that the photons from such a system can be sufficiently indistinguishable to allow for the establishment of remote entanglement. 
Here we present experimental and theoretical results of photon distinguishability in a telecom-converted ion-cavity setting, based on interference between two photons produced sequentially from an ion in a cavity. 
We conclude that the achieved interference visibilities and overall detection efficiencies would already allow for entanglement of ions tens of kilometers apart (orders of magnitude further than the state of the art \cite{Moehring2007, Hucul:2015wo, balance}).

The extent to which photons are in identical pure states, and therefore indistinguishable, is a key parameter that can be quantified by the visibility in a two-photon interference experiment \cite{PhysRevLett.59.2044}. 
For a comprehensive theoretical analysis of two-photon interference from quantum emitters without conversion see, e.g., \cite{Fischer_2016, PhysRevA.96.023861}. 
While direct two-photon interference has been achieved using neutral atoms in cavities \cite{PhysRevLett.93.070503, PhysRevLett.98.063601}, it has not previously been reported for ions in cavities. As will be shown, the limiting factor on the interference visibility in our ion-cavity system is unwanted spontaneous emission from the ion during the cavity-mediated photon generation process. Such spontaneous emission is particularly relevant for ion-cavity systems demonstrated to date in which the ion-cavity coupling rate does not overwhelm the spontaneous scattering rate. Furthermore, photon conversion stages can easily introduce additional distinguishability, e.g., by directly adding noise photons at a rate that depends strongly on the particular photon and pump laser wavelengths and filtering bandwidth \cite{Pelc:11}, and must be assessed on a system-dependent basis. 

In this Letter, first we introduce the experimental system and a simple theoretical model of the effect of spontaneous emission on the emitted cavity photon. 
Second, two-photon interference results of cavity photons at the ion-resonant wavelength are presented, showing that spontaneous emission is the dominant limiting factor. 
Third, two-photon interference results are presented after a two-step frequency conversion, converting the wavelength of one cavity photon to the telecom band and back to the ion-resonant wavelength, showing that the photon indistinguishability is essentially preserved.

Experiments employ a single $^{40}$Ca$^{+}$ atom in the center of a linear Paul trap and in the focus of a near-concentric optical cavity near-resonant with the 854 nm electronic dipole transition  (Figure 1) \cite{50km}. 
We begin by Doppler cooling the ion's motional state and optical pumping into an electronic ground state $\ket{S}=\ket{4^{2} S_{J{=}1/2},m_j{=}1/2}$ (Figure 1). 
Each photon is generated via a Raman laser pulse at  393 ~nm  which triggers  emission,  by  the  ion, of a polarized 854 nm photon into a vacuum cavity mode, via a cavity-mediated Raman transition \cite{Keller:2004cf}. 
Two photons are generated sequentially with a time gap between the beginning of their respective Raman pulses of $13.35~\mu$s, such that after delay of the first (the vertical `long-path' photon, $\ket{V}$) in a 3~km optical fibre spool, both photon wavepackets (the second being the horizontal `short-path' photon, $\ket{H}$) arrive simultaneously and with their polarizations rotated to be parallel, at different input ports of a 50:50 beamsplitter. 
Different photon polarizations are generated and modelled by 3-level Raman transitions that differ in the final Zeeman state of the $\ket{ 3^{2}D_{J{=}5/2}}$ manifold  (Figure 1 and \cite{SuppMat}). 
Specifically, after the generation of a $\ket{V}$ (an $\ket{H}$) photon the ion is in the final state $\ket{ 3^{2}D_{J{=}5/2},m_j{=}-5/2}$  ($\ket{ 3^{2}D_{J{=}5/2},m_j{=}-3/2}$)  \cite{SuppMat}. 
In every experiment we use a Raman laser Rabi frequency of $\Omega=2\pi\times 64(1)$~MHz.

\begin{figure}[th]
	\begin{centering}
		\includegraphics[width=0.9\columnwidth]{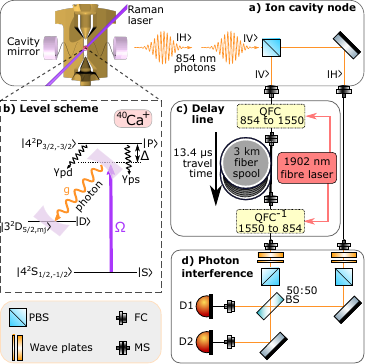}
		\caption{
			\textbf{Experiment schematic:} 
a. An atomic ion (red sphere) in a linear Paul trap (gold electrodes) and coupled to a vacuum anti-node of an optical cavity (coupling strength $g$). Raman laser (Rabi frequency $\Omega$) pulses generate sequential orthogonally-polarized photons, first vertical ($\ket{V}$) then horizontal ($\ket{H}$) that are split into two paths, with a time separation equal to the delay line (panel c.), such that their wavepackets arrive simultaneously at the beam splitter (panel d.). 
b. Three-level model: ground state $\ket{S}$; metastable state $\ket{D}$ (1.17~s lifetime) and excited state $\ket{P}$ (6.9~ns lifetime).  Spectroscopic notation shown. For $\ket{H}$ and $\ket{V}$ photons the $mj=-5/2$ and $mj=-3/2$ Zeeman states of $\ket{D}$ are used \cite{SuppMat}. 
The coherent cavity-photon generation process competes with spontaneous emission from the short-lived $P$ state (decay rates $\gamma_{ps}$ and $\gamma_{pd}$). 
c. Single photon quantum frequency conversion (QFC) and inverse process (QFC$^{-1}$) with wavelength changes shown. %
A 3~km spool of telecom SMF28-Ultra fibre. 
d. Beamsplitter (BS). Superconducting nanowire photon detector D1 (D2) with efficiency 0.88 (0.89) and free-running dark counts 0.3 (0.4) per second. Fiber coupler (FC), mating sleeves (MS). 
		}
		\label{fig:setup}
		\vspace{-3mm}
\end{centering}
\end{figure}

The arrival times of photons at the beamsplitter output ports are recorded with single-photon detectors. 
In each experimental cycle, we generate two pairs of photons: while the temporal wavepackets of the first pair (synchronous) arrive simultaneously at the beamsplitter, a time gap is introduced between the wavepackets of the second pair (asynchronous) that provides complete temporal distinguishability. 
Each full experiment consists of many repeated cycles as described in \cite{SuppMat}. 
The coincidence rates of detection events from the synchronous and asynchronous photon pairs are denoted as $C^{||}$ and $C^{\perp}$, respectively. 
The two-photon interference visibility is given by $V(T){=}1-C^{||}(T)/C^{\perp}(T)$, where $T$ is the coincidence window: the maximum time difference between photon clicks that is counted as a coincidence. 

In the first full experiment, the $\ket{V}$ photon is sent directly to the fiber spool. In the second experiment, the $\ket{V}$ photon is first converted to 1550~nm (telecom C band) via difference frequency generation (DFG) in a ridge-waveguide-integrated PPLN crystal with a 1902~nm pump laser.  This first `down-conversion' stage is described in \cite{50km, Krutyanskiy2017}. After the spool, an `up-conversion' stage (not previously reported) converts the photon back to 854~nm via the reverse process: sum frequency generation (SFG). Approximately 0.2~W of pump laser power is used for each stage. 

In the case of perfectly indistinguishable photons entering separate ports of a symmetric beamsplitter, the well known photon bunching effect occurs: two perfect detectors placed at the output ports of the beamsplitter never fire simultaneously. Yet in practice, perfect bunching is never observed, and it is important to understand the source of the imperfections.
During the photon generation process (Figure 1), spontaneous decay events from the short-lived excited state ($\ket{P}$) onto the final state manifold ($\ket{D}$) act only as losses  -- no cavity photon is emitted through the Raman process if such an event occurs. 
In contrast, following any number of spontaneous decay events from $\ket{P}$ back to the initial state ($\ket{S}$) during the Raman laser pulse, a cavity photon can still be subsequently generated while the Raman laser remains on. 
Every spontaneously scattered photon carries away the information that the cavity photon has not yet been emitted.  
Consequently, the cavity photons impinging on the beamsplitter are each in a (temporally-) mixed state and therefore they do not bunch perfectly.

The effect of spontaneous scattering on the visibility is precisely quantified through a theoretical model describing the evolution of a three-level atom embedded in a cavity using a master equation  \cite{SuppMat}. In the model, an expression for the mixed state of photons emitted from the cavity is obtained in two steps. First, we calculate the wave function of photons emitted from the cavity conditioned on the ion being in the initial state $\ket{S}$ at time $s$ and no spontaneous decay events happening for later times. Second, we compute the rate of spontaneous decay events from $\ket{P}$ to $\ket{S}$ as a function of time. The state of the emitted cavity photon is then expressed as a mixture over all the possibilities where the last $\ket{P}{\rightarrow}\ket{S}$ decay happens at time $s$ or no decay events occur and a pure state photon is emitted afterwards, plus the vacuum component collecting all the possibilities where no cavity photon is emitted. With the emitted photon states in hand, it is then straightforward to calculate the visibility of pairs of photons \cite{SuppMat}. We refer to this model that includes only imperfections due to spontaneous scattering as the basic model. 
As an alternative from our model, the visibility could be computed from the master equation via the quantum regression theorem \cite{Fischer_2016}. 

\begin{figure}[t]
	
	\begin{centering}
		\includegraphics[width=1\columnwidth]{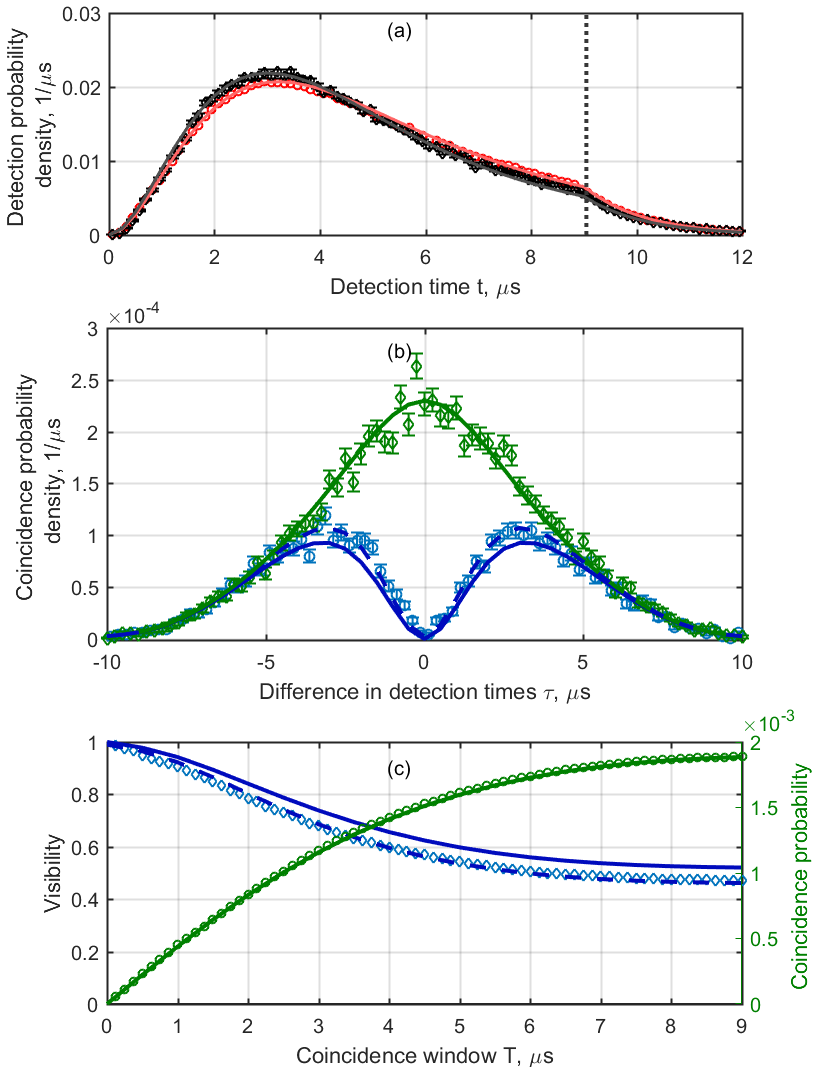}
			\caption{
			\textbf{Two-photon interference without photon conversion:}
			Solid (dashed) lines show basic theory (extended theory) model and shapes show data in all the panels. Probability densities are obtained by dividing the probability of detection (coincidence) per time bin by the bin size, see \cite{SuppMat} sec. IV, V for details. 
			(a) Single photon wavepackets for short path (red 
			circles) and long path (black diamonds, rescaled by multiplication factor 4.6 to correct delay line losses). Vertical dotted line shows the end of the Raman laser pulse. 
			(b) Photon coincidences for temporally synchronous ($\rho_{c}^{||}$, blue circles) and asynchronous 
			($\rho_{c}^{\perp}$, green diamonds) cases.
			(c) Interference visibility $V$ (left axis, blue diamonds) and integrated asynchronous coincidence probability $C^{\perp}$ (green circles, right axis). Error bars represent  $\pm$ one standard deviation due to Poissonian photon counting statistics, not shown when smaller than shapes. 
		}
		\label{figure_results_854}
	\end{centering}
\end{figure}

Results are now presented for the case without photon conversion. 
The temporal profiles of the short-path and long-path single photon detection events from the second (asynchronous) photon pair are shown in Figure \ref{figure_results_854}a. These single-photon wavepackets are presented as a probability density $\rho_{d}(t) = N_d/(k\cdot \Delta_t)$, where $N_{d}$ is the number of detection events registered in a time bin $\Delta_t = 125~$ns and $k$ is the number of trials. 
Integration of  the wavepackets gives the probability of detecting a short- (long-) path photon as $12.4$\% ($2.7$\%).
Differences in the single-photon wave packet shapes are due to slight differences in the corresponding transition strengths \cite{SuppMat}.

The temporal profile of the coincidence detection events (cross-correlation function) for the synchronous and asynchronous photon pairs are compared in Figure \ref{figure_results_854}b. Here the coincidence probability density $\rho_{c}(\tau)^{\parallel,\perp} = N^{\parallel,\perp}_c/(k\cdot \Delta_\tau)$ is used, where $N^{\parallel,\perp}_{c}$ are the number of coincident detection events per time bin for the first and second pair of photons respectively, $\tau$ is the difference in detection times. 
Figure \ref{figure_results_854}c shows the visibility $V(T)$  and integrated coincidence rate of the asynchronous photons $C^\perp(T) = \int_{-T}^T \rho^\perp_c(\tau)d\tau$.
For the smallest coincidence window presented, the interference visibility $V($125~ns) is $0.986\pm{0.006}$ ($0.987\pm{0.005}$ after subtracting detector dark counts). 
When considering a coincidence window containing the whole photon wavepacket, the visibility $V(9~\mu$s) is $0.472\pm{0.008}$. 
From the theory, we calculate that the expectation value of the number of spontaneously-emitted photons on the $\ket{P}{\rightarrow}\ket{S}$ transition, given generation of a cavity-photon, was 3.5. 

The differences between the basic model and data in Figure \ref{figure_results_854} are consistent with an extension to the model that, in addition to spontaneous emission, includes a combination of an overall time-independent distinguishability factor of 1\% and a centre frequency difference of the two photons of 40~kHz \cite{SuppMat}. This  small photon frequency difference could be caused by several reasons, e.g., cavity length instability, acoustic noise in the 3 km delay line fibre and cavity birefringence.
The 1\% time-independent distinguishability can arise from slight polarization mode mismatch at the beamsplitter or imbalance of the 50:50 beamsplitter itself.  
The agreement between data and the basic model shows that we are close to the fundamental limit of photon indistinguishability set by spontaneous scattering in our system. 

\begin{figure}[t]
	
	\begin{centering}
		\includegraphics[width=1\columnwidth]{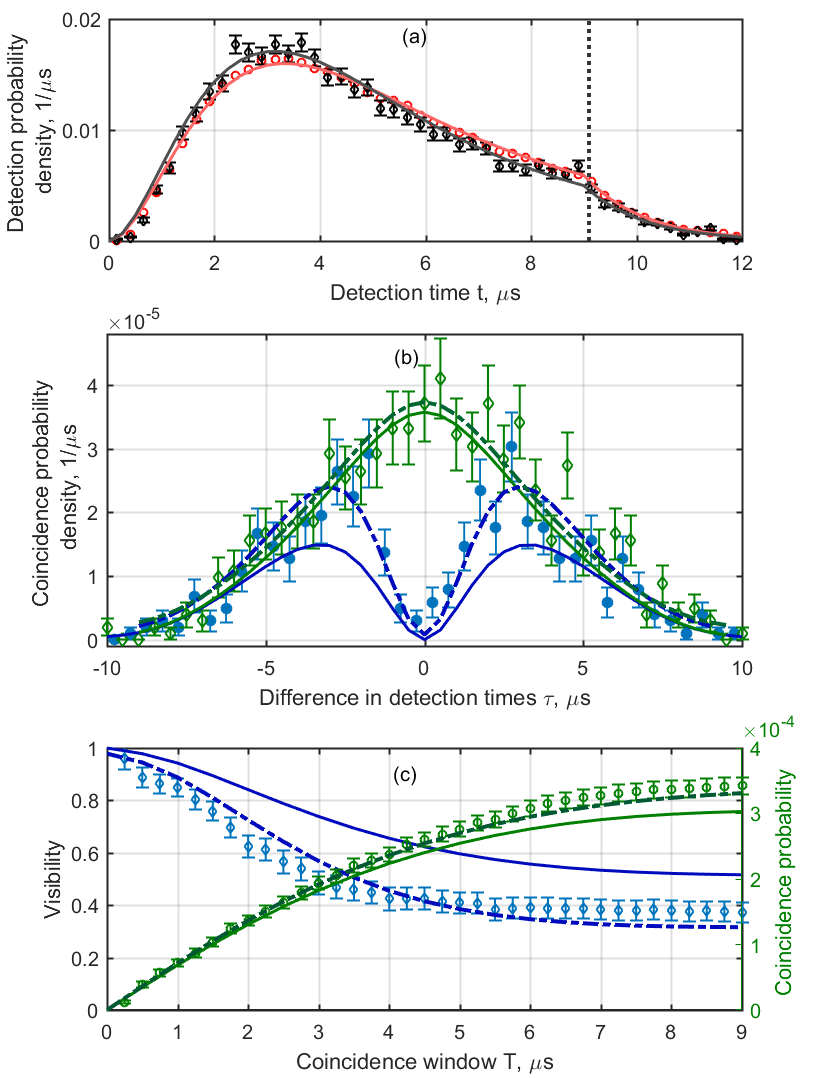}
		\caption{
			\textbf{Two-photon interference with photon conversion:}
			Solid (dotted) lines show basic theory (extended theory) model and shapes show data in all the panels. Probability densities are obtained by dividing the probability of detection (coincidence) per time bin by the bin size \cite{SuppMat}. 
			(a) Single photon wavepackets for short path (red 
			circles) and long path (black diamonds, rescaled by multiplication factor 19 to correct delay line losses). Vertical dotted line shows the end of the Raman laser pulse.  
			(b) Photon coincidences for temporally synchronous ($\rho_{c}^{||}$, blue circles) and asynchronous
			($\rho_{c}^{\perp}$, green diamonds) cases.
			(c) Interference visibility $V$ (left axis, blue diamonds) and integrated probability $C^{\perp}$ (green circles, right axis). Error bars represent  $\pm$ one standard deviation due to Poissonian photon counting statistics. 
		}
		\label{figure_results_converted}
	\end{centering}
\end{figure}

Figure \ref{figure_results_converted} presents results for the case with photon conversion and is constructed in the same way as Figure \ref{figure_results_854}. 
The probability of detecting a short- (long-) path photon across the entire wavepacket is $10$~\% ($0.5$~\%).
The visibility is 0.96$\pm{0.04}$ for the minimum coincidence window of 250~ns and 0.37$\pm{0.04}$ for the full wave packet window of 9 $\mu$s. 
The differences between the basic model and frequency-converted data (Figure \ref{figure_results_converted}) are consistent with an extended model that includes a frequency drift of the unstabilized photon conversion pump laser at the level of 50~kHz on a 10$~\mu$s timescale (consistent with independent measurements) and background coincidences, see section V.C \cite{SuppMat}. 
We anticipate no significant challenges to frequency stabilising the pump laser to the few kilohertz level in future work. Remote photon conversion stages in distributed networks will need independent pump lasers with absolute long-term frequency stability to within a fraction of the networking photon bandwidth.

The achieved visibilities and coincidence rates in our experiments would already allow for remote ion entanglement over tens of kilometers. 
Consider the entanglement swapping protocol of \cite{Duan2010}, which leads to maximal entanglement of two remote (ion) qubits with state fidelity $F(T){=}(1+V(T))/2$ \cite{PhysRevLett.123.213601} at a heralded rate $R_{swap}(T) \propto R_{gen} \times C^{\perp}(T)$, where $R_{gen}$ is the photon-generation attempt rate at each ion-trap network node. 
Using $R_{gen}=30$~kHz, the achieved values without photon conversion (Figure \ref{figure_results_854}) would allow for  3~km ion-ion entanglement distribution with $F(T=9\mu s)=0.736 \pm0.004$ 
at a rate of $R_{swap}(9\mu s)=30$~Hz.  
Using 0.18 dB/km for telecom fiber losses, the achieved performance with photon conversion (Figure \ref{figure_results_converted}) would allow for 50~km distant ion-ion entanglement generation with $F(T=9\mu s)=0.69 \pm0.02$ 
at rates on the order of 1~Hz (assuming photon detector dark count rates of 1~Hz).
In all cases, time filtering of the coincidences ($T<9\mu$s) would allow for an increased remote entangled-state fidelity at the cost of reduced heralding rate. 
Such long-distance experiments must tackle environmental noise in deployed optical fibres, absolute frequency stabilisation of remote laser systems and matching photons from remote network nodes. 
Methods to improve visibility without reducing the photon generation rate are those that can significantly increase the coherent ion-cavity coupling rate $g$, such as coupling multiple ions in entangled (superradient) states to the cavity \cite{PhysRevLett.107.030501, PhysRevLett.114.023602} and pursuing small mode-volume fibre cavities \cite{PhysRevLett.110.043003}.

Our model reveals that there is an optimal drive-laser Rabi frequency ($\Omega$, Figure 1) that achieves the highest photon generation probability (and therefore $C^{\perp}$) for a given threshold visibility, 
highlighting the important role such models will play in enabling the upcoming next generation of long-distance networking experiments \cite{SuppMat}. 
Our results present a path to distributing entanglement between trapped-ion registers spaced by several tens of kilometres at practical rates for verification: significantly further than state-of-the-art experiments involving spacings of a few meters \cite{Moehring2007, Hucul:2015wo, balance} and a practical distance to start building large-scale quantum-logic-capable quantum networks. \\

\noindent \emph{Note:} During the preparation of this manuscript, we became aware of complementary work in which sequential interference of photons from an ion in a cavity is achieved and studied \cite{walker2019improving}.

\begin{acknowledgments}

This work was supported by the START prize of the Austrian FWF project Y 849-N20, by the US Army Research Laboratory under Cooperative Agreement Number W911NF-15-2-0060 (project SciNet), by the Institute for Quantum Optics and Quantum Information (IQOQI) of the Austrian Academy Of Sciences (OEAW), by the European Union's Horizon 2020 research and innovation programme under grant agreement No 820445 and project name `Quantum Internet Alliance' and by the Swiss National Science Foundation (SNSF) through the Grant PP00P2-179109, and by the Austrian Science Fund (FWF) through Project F 7109. The European Commission is not responsible for any use that may be made of the information this paper contains.

\section*{Author Contributions}

VKrut., MM, and VKrc. took the data. 
VKrut, AM, BPL, PS and MM analysed the data.
VKrut, MM, VKrc, JS and BPL contributed to the experimental setup and design. 
AM, VKrut., PS, BPL, MM, DF, TN and NS performed theoretical modelling. 
BPL, VKrut, MM, PS, TN and NS wrote the majority of the paper, with contributions from all authors. 
The project was conceived and supervised by BPL.
		
\end{acknowledgments}


%

\hypertarget{sec:appendix}
\appendix
\setcounter{section}{0}
\setcounter{subsection}{0}
\renewcommand{\thesubsubsection}{\arabic{subsubsection}}
\renewcommand{\thesubsection}{\Alph{subsection}}
\renewcommand{\thesection}{\Roman{section}}
\renewcommand{\thefigure}{A.\arabic{figure}}
\setcounter{equation}{0}
\numberwithin{equation}{section}

\newpage

\section*{Supplementary Material}
 \tableofcontents

\newpage

\section{Experimental setup}

\begin{figure*}
	\vspace{0mm}
	\begin{center}
		\includegraphics[width=1\textwidth]{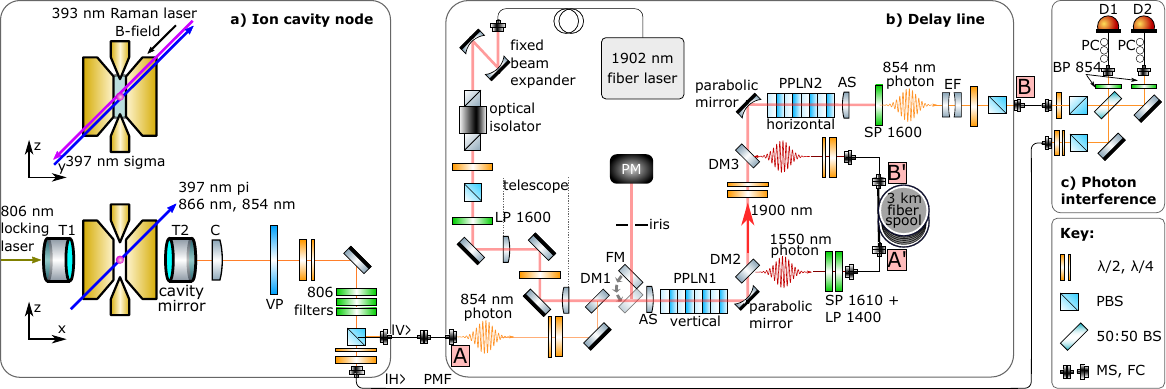}
		\vspace{0mm}
		\caption{
			\textbf{Detailed experimental diagram.} 
			\textbf{a) Ion cavity node.}
			A single atomic ion (red sphere) in the centre of both a 3D radio-frequency (RF) linear Paul trap (gold electrodes) and an optical cavity.  The two smaller electrodes are held at DC voltage.
			The 4 larger electrodes (two shown in figure projection) are driven with RF. Two cross sections are depicted: Along the cavity axis (top), showing: the $\approx4$ Gauss DC magnetic field (quantisation axis) generated by rings of permanent magnets and the circularly-polarized Raman laser for generating 854 nm cavity photons.
			Following a Raman pulse, an 854 nm cavity photon exits the cavity via the right mirror (transmission T2). The photon then passes the following elements: in-vacuum collimating lens (C); vacuum chamber viewport (VP); waveplates; 3 filters to remove the 806 nm laser light to which cavity length is continuously and actively stabilised; polarizing beam splitter (PBS) for directing the vertical photons into a single mode fiber, and the horizontal photons into a polarization maintaining fiber (PMF),
			\textbf{b) Delay line.}
			For the experiment without photon conversion, point A and A', as well as the points B' and B directly are fiber connected , such that the vertical photon only has to pass the 3 km fiber spool (Corning SMF-28 Ultra).
			For the experiment with the conversion, the setup is as shown here.
			The injected 854 nm photon passes waveplates (used for system optimization with classical light) and is overlapped with 800 mW of 1902 nm laser light (Tm-doped fiber laser, AdValue Photonics AP-SF1-1901.4-01-LP, measured at PM) on a dichroic mirror (DM1) and free-space coupled into one of the ridge waveguides of temperature-stabilised PPLN1 using an asphere (AS, 11 mm, positioned by an XYZ translation stage).
			The 1902 nm input path is described in \cite{50km}.
			A gold parabolic mirror (f = 15 mm) is used to collimate all fields at the output of PPLN1.
			A dichroic mirror (DM2; Thorlabs DMLP1800) splits the converted 1550 nm photons from the 1902 nm pump laser. A combination of a shortpass (SP 1610) and longpass (LP 1400) filter reduces unwanted pump laser and other noise light fields.
			The 1550 nm photon couples into the 3 km SMF-28 fiber spool, which is used as an optical delay line. The output of the fiber spool passes waveplates, to correct for polarization rotations through the fiber and is overlapped (with a dichroic mirror DM3; Thorlabs DMLP1800) back with the 1900 nm pump light, which passes waveplates to set correct pump power for the second crystal (PPLN2).
			Via a second gold parabolic mirror (f = 15 mm) all fields are coupled into the second temperature controlled chip PPLN2, where the 1550 nm photon is converted back to the initial 854 nm via the reverse (upconversion) process. An Asphere (AS) collimates the output field, before a shortpass (SP 1600; OD5 at 1902) filters the 1902 nm pump laser from the 854 nm single photons.
			After passing an etalon filter (EF; LightMachinery, Bandwidth $\approx 870~$MHz) a combination a waveplate and PBS is used to filter unpolarized noise photons before coupling the 854 nm photon into a single mode fiber, which goes to the interference board.
			\textbf{c) Photon interference.}
			Both inputs pass waveplates and PBSs (cleaning polarization) and overlap on a 50:50 beamsplitter (50:50 BS). Both outputs of the beamsplitter are filtered with an 854 nm bandpass filter (BP 854); coupled to single mode fibers; polarization control paddles (PC) correct to most efficient polarization for followed single photon detector (D1(D2): Scontel, efficiency 88\%(87\%), dark count rate 0.5 s\textsuperscript{-1}(0.3 s\textsuperscript{-1}). The electronic pulses produced by the detectors are detected with a time tagging module (Swabian Instruments Time Tagger 20). Mating sleeves (MS), Fiber coupler (FC).
		}
		\label{fig:setup_full}
		\vspace{-6mm}
	\end{center}
\end{figure*}

\noindent A detailed experimental diagram is presented in Fig. \ref{fig:setup_full}. 
For details on the ion trap and cavity system please see our recent paper in \cite{50km}: the same ion trapping frequencies, geometry of laser beams and optical cavity were employed. A few key parameters are now recapped. 

The optical cavity around the ion is near-concentric with a cavity waist of $\omega_0 = 12.31 \pm 0.07$ $\mu$m and a maximum ion-cavity coupling rate of $g_{0} = 2\pi \cdot 1.53 \pm 0.01$ MHz. 
The finesse of the cavity (at 854 nm) is $\mathcal{F} = \frac{2\pi}{\mathcal{L}} = 54000 \pm 1000$, with the total cavity losses $\mathcal{L} = T_1 + T_2 + L_{1+2} = 116 \pm 2 $ ppm, determined from measurements of the cavity ringdown time.
This gives the cavity linewidth $2\kappa = 2\pi \cdot 140 \pm 3$ kHz, $\kappa$ being the half-width at half maximum. 

As explained in the main text, in the second experiment the delayed photon undergoes a two stage frequency conversion. 
Each frequency conversion uses a single 48 mm-long ridge-waveguide-integrated PPLN crystal (NTT electronics, reported in \cite{Krutyanskiy2017, 50km}) where the input single photon is overlapped with the pump laser at 1902 nm.  
The first (`down-conversion') stage brings the 854 nm (V) photon to the 1550 nm telecom C-Band ($(854~\mathrm{nm})^{-1}-(1902~\mathrm{nm})^{-1}\approx(1550~\mathrm{nm})^{-1}$) as detailed in the first experiment in \cite{Krutyanskiy2017}. 
This photon is separated from the pump field with a dichroic plate, injected into the 3 km spool and recombined with the pump before the second stage. 
The second (`up-conversion') stage brings the 1550 nm photon back to the original 854 nm wavelength via sum frequency generation (SFG). 
The total pump laser power of correct polarization for conversion in-coupled into each crystal is 0.2~W, control by the waveplate angles before and after the first stage.

While in the ideal case the second stage is the reverse process of the first, there are two differences which are now summarized.
In the downconversion process, unwanted photons at the telecom output wavelength are generated directly from Anti-Stokes Raman scattering (ASR) of the pump laser from 1902~nm to 1550~nm.
In the upconversion process, unwanted photons at the 854~nm output wavelength are generated by a two stage process. First, as before ASR scattering of the pump laser generates telecom photons. Second, these photons are up-converted back to 854nm via SFG \cite{Pelc:11, Kaiser15}. 
A second difference is the fundamental waveguide mode at 1550~nm has a higher overlap with the free-space Gaussian mode of the in-coupling laser field, than the waveguide mode at 854~nm. As such, we achieve a higher in-coupling efficiently into the fundamental waveguide mode at 1550~nm and subsequently a higher conversion efficiency for the up-conversion (presented in the next section).

\section{Photon conversion performance}

\begin{figure*}
	\vspace{0mm}
	\begin{center}
		\includegraphics[width=0.9\columnwidth]{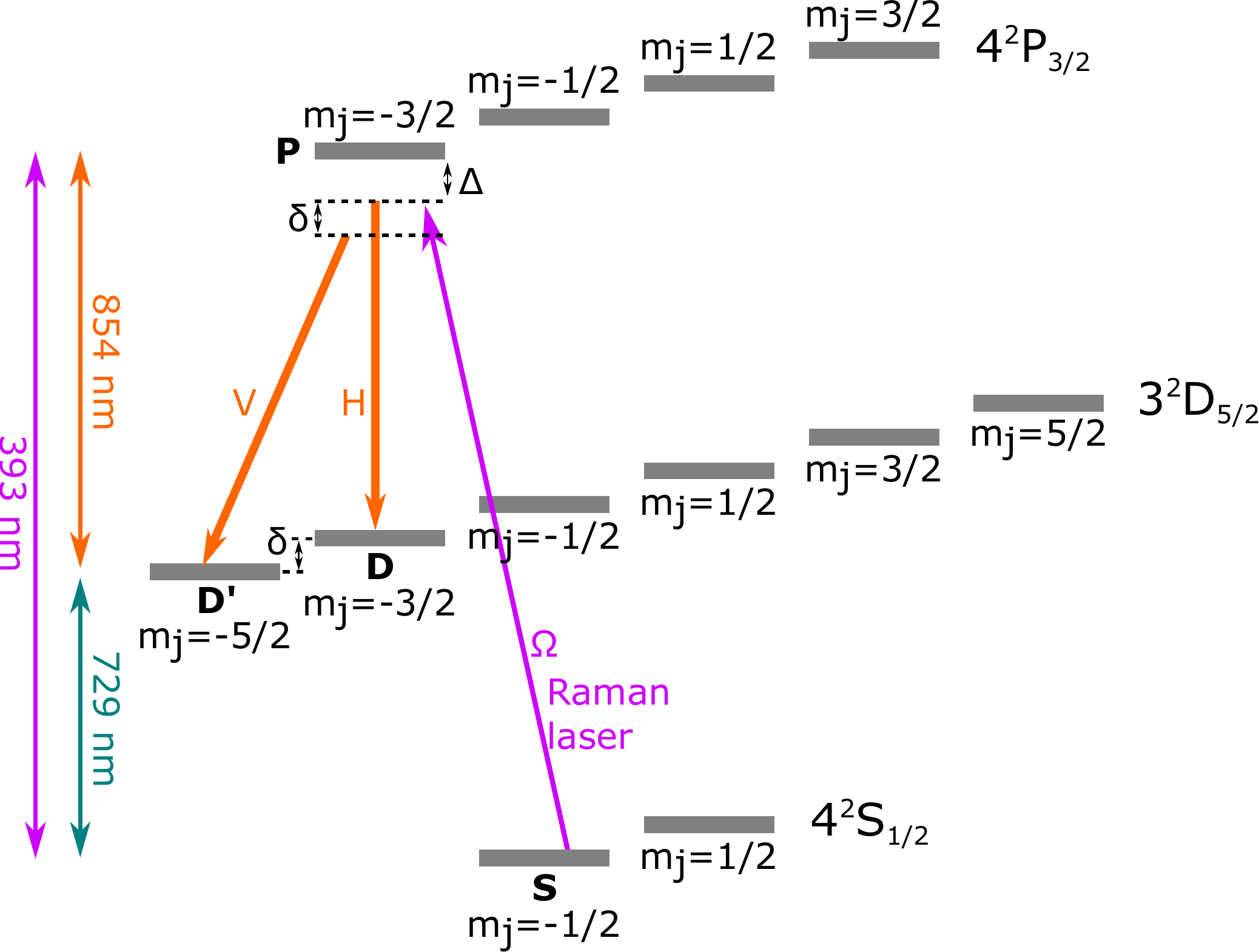}
		\vspace{0mm}
		\caption{\textbf{Relevant energy level scheme of $^{40}$Ca$^+$.}  
			Following optical pumping, the ion (single outer valence electron) begins in the state $4^2$S$_{1/2,m_j=-1/2}$.  Photons are generated via a cavity-mediated Raman transitions (CMRT) \cite{Keller:2004cf, Barros_2009, Stute2012, 50km}. 
			In case of detuning $\delta=0$, a 393~nm Raman laser pulse leads to the generation of a horizontally-polarized $\ket{H}$ photon in the optical cavity and the electron ends up in the state $3^2$D$_{5/2,m_j=-3/2}$. In the case of a different appropriate detuning $\delta$ to match the D-state Zeeman splitting (achieved by change of frequency of the Raman laser) a vertically-polarized $\ket{V}$ photon in generated in the optical cavity.  Here H stands for the linear polarization along the B-field quantisation axis  ($\pi$-photons). The V polarization stands for the orthogonal linear polarization, produced by the projection of the $\sigma$-polarized photons into the plane perpendicular to the cavity direction (and perpendicular to the B-field quantisation axis). The overall detuning $\Delta=403\pm 5$ MHz.}
		\label{fig_levelscheme}
		\vspace{2mm}
	\end{center}
\end{figure*}

\noindent A set of characterisation measurements of the two-stage conversion process was carried out using classical laser light and the results are now described. 
The total efficiency from the start of the converted delay line (Point A in figure A.1), to the end of the delay line (point B in figure A.1) was measured to be $0.098\pm 0.005$ at the last calibration before the two-photon experiments (the error stands for the last digit of the powermeter reading). 

The efficiencies of separate parts were characterized independently. The values below stand for the final characterisation after alignment-optimisation before the two-photon experiment reported in the main text. The intervals are derived from the calibration measurements after optimisation on other days, representing maximum and minimum values observed. The laser powers and setup stability on the timescale of performing the characterisation is also taken into account. 
The down-conversion stage efficiency from the point A to the in-coupling of the delay fiber was $0.50\pm 0.03$ (we refer it as down-conversion external efficiency). 
The delay fiber transmission was measured to be $0.6\substack{+0.01 \\ -0.05}$, including in-coupling ($0.75\substack{+0.005 \\ -0.05}$) and two mating sleeves ($0.95\pm 0.02$).
The up-conversion external efficiency (from the delay fiber out-coupler to the etalon) was measured to be $0.53\pm 0.03$. 
The etalon transmission was $0.84\substack{+0.005 \\ -0.04}$ and the coupling to the fiber that goes to the HOM board was $0.73\substack{+0.05 \\ -0.03}$. 
The provided value above of external up-conversion efficiency of  $0.53\pm 0.03$ includes the wavegide in- and out-coupling losses and transmission of the filter that blocks the pump field (see \rfig{fig:setup_full}).  

We define the internal conversion efficiency as conversion efficiency without coupling and transmission losses, calculated as the fraction of converted signal-photon-number to unconverted signal-photon-number (without pump light) at the output of the converter. 
We observe at best alignment 89(0.5)\% internal efficiency for the upconversion and estimate a waveguide coupling/propagation efficiency of 70(5)\%, limited by the mismatch of the out-coupler of the delay fiber and the in-coupler of the conversion waveguide. For the down-conversion, the performance was reported in \cite{Krutyanskiy2017} and yields 66(6)\% internal efficiency if the waveguide in-coupling/propagation losses are taken into account according to the definition above. This efficiency was shown to be limited by the unintentional excitation of higher-order waveguide modes.
In order to prove that the up-conversion coupling/propagation efficiency of 70(5)\% is dominated by coupling, we performed a separate measurement with adjusted 1550 beam diameter before  the wavguide in-coupler and achieved 82(3)\% efficiency calculated as the ratio of the number of photons at 854 nm wavelength right after the waveguide to the number of 1550 nm photons right before the waveguide. Note, that the external and internal up-conversion efficiencies reported here are higher than the ones achieved in similar systems before \cite{Maring2017, Kaiser:19}.

On the day of the two-photon interference experiment using photon conversion (Figure 3, main text), the total efficiency of the delay line was measured to be approximately 5\%. This lower efficiency, compared to the aforementioned values achieved with classical light is caused by the combination of the following: an additional fiber joiner between the ion node and delay line (panels a and b in Figure \ref{fig:setup_full}) when working with single photons (compared to classical light); imperfectly optimised fiber couplers throughout the delay line; and potential slight mismatch between the photon polarization and the non-linear crystal axis.

The noise, at the single photon level, introduced by the conversion process is now presented. These values are extracted from the photon detector click rate outside of the known ion-photon arrival times recorded in the experiment presented in Figure 3 of the main text. Recall that final narrowband filtering is performed at 854 nm via a temperature-controlled etalon (Bandwidth 870 MHz, free spectral range 30 GHz) that has a maximum transmission of 84\%. 
From the measured photon noise rate of $11\pm3$ cps at the detectors (after removing detector dark counts and other known background noise, like room light) we estimated from the known losses in the optical path between the etalon and the detectors a noise level of $50\pm 10~\text{s}^{-1}$ right after the final etalon filtering stage.

\section{Photon generation sequence}
\begin{figure*}
	\vspace{0mm}
	\begin{center}
		\includegraphics[width=\textwidth]{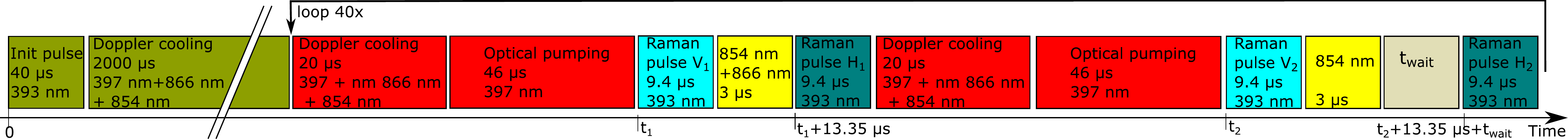}
		\vspace{0mm}
		\caption{
			\textbf{Sequence of the laser pulses during the experiment.} Wavelengths and duration of pulses are labeled. Each cycle (what is shown within the loop) contains four Raman laser pulses, which attempt to generated four photons. The first two are referred to as the synchronous pair and the second pair as the asynchronous pair. Each cycle is looped 40 times and this sequence is repeated thousands of times to produce the data presented in the main text. 
		}
		\label{fig:sequence}
		\vspace{-6mm}
	\end{center}
\end{figure*}

We sequentially generate photons of orthogonal polarizations using a cavity-mediated Raman transition (see \rfig{fig_levelscheme}) with ion state reinitialisation in between. The full experimental sequence is shown in \rfig{fig:sequence}. First, a $40~ \rm{\mu s}$ `initialisation' laser-pulse at 393 nm is measured by a photodiode in transmission of the ion-trap chamber and used for intensity stabilisation of the subsequent 393~nm photon-generation Raman pulses with a sample-and-hold system. The initialisation pulse is followed by $2000~ \rm{\mu s}$ of Doppler cooling, involving three laser fields as indicated.
Next, a cycle starts in which the photon-pair generation attempt takes place. This cycle is repeated (looped) 40 times before the whole sequence starts again.

In summary, each cycle contains four Raman pulses --- \Vo, \Ho~and \Vt, \Ht~--- which attempt to generate the two pairs photons that are refereed to as `synchronous' and `asynchronous' in the main text. The first synchronous pulse pair (\Vo, \Ho) has a time difference of 13.35~$\mu$s, corresponding to the length of the delay line, such that the generated  photon wavepackets arrive at the interference beamsplitter simultaneously (the delay was measured with $<50$ ns accuracy by recording the photon arrival times). 
The second asynchronous pulse pair (\Vt, \Ht) has an additional delay $t_{wait}$ and generates a fully temporally distinguishable photon pair as a reference ($t_{wait} = 30 \mu$s). Before the \Vo~pulse in each loop we produce an electronic trigger-pulse that is recorded on a separate channel of the time-tagger, along with the photon detection events, to provide exact Raman pulses timing information.

In detail, each cycle starts with an additional Doppler cooling pulse ($20 ~\rm{\mu s}$) and optical pumping to the $S_{J=1/2,m_j=-1/2}$ (see \rfig{fig_levelscheme}) state via circularly polarized 397~nm laser light ($46 ~\rm{\mu s}$).
The photon-generation Raman pulse \Vo ($9.4 ~\rm{\mu s}$) creates the vertical polarized photon that is directed to the delay line by a PBS.
This is followed by a $4~ \rm{\mu s}$ long, $854 ~\rm{n m}$ repump pulse which pumps the ion back to the initial ground state.
A second photon-generation Raman pulse \Ho ($9.4 ~\rm{\mu s}$) creates a horizontal photon that is directed directly to the interference region. 
After a Doppler cooling pulse of $20 ~\rm{\mu s}$ and optical pumping of $46 ~\rm{\mu s}$, the second pair of photons is produced.

\section{Data analysis}

During the experimental run we record the absolute time stamps of two detector events 
$D_1$ and $D_2$, and of an electronic trigger pulse generated simultaneously with \Vo~at time $t_1$ in each cycle (see \rfig{fig:sequence}). 
We then work in a time frame referenced to the trigger pulse. 
In this frame the photon arrival times are grouped into three time windows: the first group contains the overlapped synchronous photons (generated by \Vo~and \Ho), the second group contains the first of the time-displaced `asynchronous' photons 
(\Vt, through the delay line) and the third group contains the later asynchronous photon (\Ht, direct path).
In the main text Figure 2a we sum up the data for two 
detectors and plot separately the probabilities of events corresponding to (windows containing) \Vt  and \Ht. 
This is done by shifting the distributions in time by a fixed offset $t$ ($t+t_{wait}$) for 
\Vt (\Ht), with $t$ being the delay between the trigger pulse and the expected \Vt-photons' front-slope; $t_{wait}$ is the known additional wait time (30 or 40 
$\mu$s in different realisations). 
We plot the detection-probability-density (see main text), defined as the number of events detected in a certain time bin during the experiment divided by the number of trials and bin duration. 
The error bars represent $\pm 1$ standard deviation of Poissonian photon counting statistics. The subtraction of background counts was performed for a correct efficiency comparison of the two paths.

To plot the coincidence distribution (Figures 2b, 3b of the main text) we first choose a time window (software gate) where the corresponding photons are expected to arrive based on sequence timing and the histogram of all recorded events.  
Then we calculate the probability density  $\rho_c(D_1(t_1), D_2(t_2))$ of observing a two-photon detection event in a given trial as a function of the detections time difference $\tau = t_2-t_1$. 
The plotted values in the figure are calculated as $\rho_c(\tau) = \frac{1}{\Delta_t k}\int_{gate}dt_1 \int_{t_1+\tau}^{t_1+\tau+\Delta_t}dt_2 N(t_1,t_2)$ where $N(t_1,t_2)$ is the number of two-photon clicks with given times, $\Delta_t$ is the bin size in the figure and $k$ is the total number of attempts.
The errorbars for each point in the figure are calculated from the total number of events detected for this bin assuming Poissonian statistics.
The coincidence distribution for the distinguishable photons (from \Vt, ~\Ht), originally peaking at the $\tau = \pm t_{wait}$, is shifted to $\tau = 0$ and summed over positive and negative branches to represent the expected coincidence distribution for the fully distinguishable but synchronized photons. 

Given the coincidences distribution in time we define the visibility (plotted in fig. 2c, 3c of the main text) as:
\begin{equation}
V(T) = \frac{C^\perp(T)-C^\parallel(T)}{C^\perp(T)}, 
\label{eq_Visibility_exp}
\end{equation}
where $C^\perp(T)$ ($C^\parallel(T)$) are the coincidence probabilities for the distinguishable (overlapped) pair of photons plotted in Figures 2b,3b of the main text integrated over the delay range $\tau\in[-T;T]$: $C(T) = \int_{-T}^T \rho_c(\tau)d\tau$. 

For the passive delay-line experiment (without conversion) we perform in total 7.5 million cycles, where each 
cycle corresponds to one loop (cycle) in fig \ref{fig:sequence} (attempt to generate four photons). The experiment with frequency conversion consists of a total of 2.5 million cycles.  

\section{Theory model}
\subsection{The Master Equation}

\begin{figure}
	\centering
	\includegraphics[width=0.4 \textwidth]{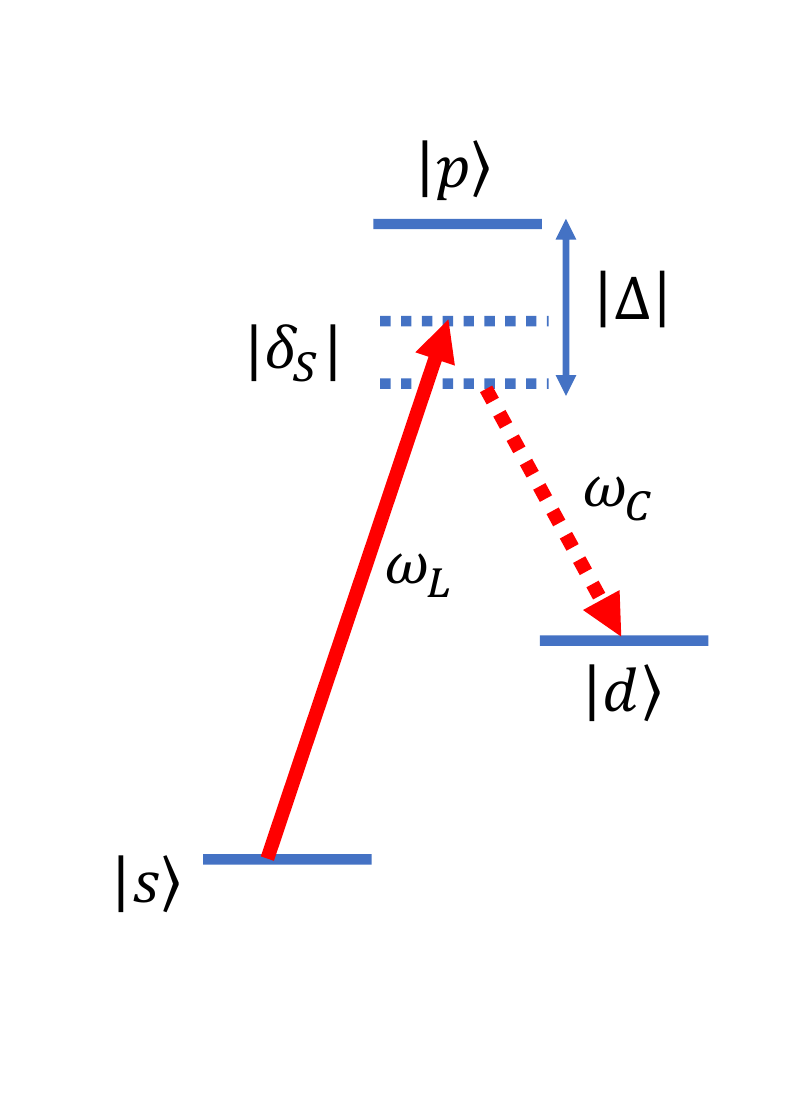}
	\caption{Scheme of the $\Lambda$-system relevant for the experiment. The transition between $|s\rangle$ and $|p\rangle$ is off-resonantly driven by a pump laser with frequency $\omega_L$ while the states $|p\rangle$ and $|d\rangle$ are coupled by the field of a cavity with frequency $\omega_C.$ The explicit expressions of each detunings are given in the text.}
	\label{fig:lambda system}
\end{figure}

We start by writing down a master equation for a single $^{40}$Ca$^+$ ion trapped inside a cavity and driven by a pump laser. We restrict the atomic model to a $\Lambda$-system formed by three levels  $\ket{s}, \ket{p}$ and $\ket{d}$ (see Fig.~\ref{fig:lambda system}) corresponding to sublevels of $S_{J=1/2,m_j-1/2},$ $P_{J=3/2,m_j=-3/2}$ and $D_{J=5/2,m_j=-5/2}$ (or $D_{J=5/2,m_j=-3/2}$) that are of direct importance for the experiment, see Fig.~A.2 for details. The ion is initially prepared in the state $\ket{s}$. The laser is driving off-resonantly the $p-s$ transition with a frequency $\omega_{L} =\omega_{ps}+ \Delta -\delta_S,$ where both $\Delta$ and $\delta_s$ are negative. We denote $a, a^\dag$ the bosonic operators associated to the cavity field whose frequency is given by $\omega_C= \omega_{pd}+\Delta.$ In the experiment, the laser Rabi frequency $\Omega_t$ is much lower than the detuning $|\Delta|$ and the state $|s\rangle$ undergoes a Stark-shift $\Omega_t^2/4\Delta.$ The additional laser detuning $\delta_S=\Omega_t^2/4\Delta$ is chosen to preserve the two-photon resonance between $s-d$. The Hamiltonian of the atom-cavity system is given by
\be\label{eq:Hamil}
\begin{split}
	H&= \omega_C a^\dagger a+\omega_{ps}\prjct{p}+\omega_{ds}\prjct{d} \\
	&+\frac{1}{2}(e^{i\omega_L t}+e^
	{-i\omega_L t})(\Omega_t\ketbra{s}{p}+\Omega_t\ketbra{p}{s}) \\
	&+ g(\ketbra{d}{p}+\ketbra{p}{d})(a^\dag+a),
\end{split}
\ee
where we have set $\hbar$ to one. The Hamiltonian can be simplified by noting that the cavity mode is initially empty $\ket{0}.$ Therefore, without additional coupling terms the atom-cavity system remains in the three level manifold \{$\ket{s,0}, \ket{d,1}, \ket{p,0}$\}. Under the rotating wave approximation, the Hamiltonian in this subspace is thus given by
\be
H_t= \left(\begin{array}{ccc}
	0&0 & \Omega_t/2  \\
	0& \delta_S &g\\
	\Omega_t/2& g & -\Delta + \delta_S
\end{array}
\right),
\ee
in the rotating frame with $a^\dag \to e^{-\ii\omega_C t}a^\dag$, $\ket{p}\to e^{-\ii \omega_L t}\ket{p}$, $\ket{d}\to e^{-\ii (\omega_L-\omega_C)t}\ket{d}$, and $\ket{s}\to \ket{s}$.  \\

Let us now introduce the non-unitary terms which will appear in the master equation. First, the photon can escape the cavity mode with rate $\kappa$. This decay channel is precisely the one in which the photons are collected into a fiber and sent to the detectors. Yet, for the atom-cavity system, this process corresponds to a loss term $L_1=\sqrt{2\kappa} \ketbra{d,0}{d,1}.$ Note that in practice, photons can leave the cavity through channels that are not the detected channel. This additional cavity loss is taking into account by adjusting the photon detection efficiency. 
Finally, there are two scattering terms $L_2= \sqrt{2\gamma_{sp}} \ketbra{s,0}{p,0}$ and $L_3= \sqrt{2 \gamma_{dp}}\ketbra{d,0}{p,0}$. With this in hand, we can write down the master equation for the atom-cavity system
\be\label{eq:ME}
\dot \varrho_t = -\ii\, [H_{t},\varrho_t] + \sum_{i=1}^3 \left(L_i \varrho_t L_i^\dag - \frac{1}{2}\{L_i^\dag L_i, \varrho_t\}\right),
\ee
where $\varrho_t$ has to be defined on a four level manifold including $\ket{d,0}.$ $H_{t}$ is extended trivially on the added level via $H_{t} \ket{d,0}=\delta_{s} \ket{d,0},$ that is, $\ket{d,0}$ is not coupled to the other three levels. Hence, the atom-cavity system evolves in the $\{\ket{s,0},\ket{d,1},\ket{p,0}\}$-manifold until it is brought to the state $\ket{d,0}$ either by the scattering term $L_3$ or by emitting a photon towards the detector via $L_1.$ \\

\subsection{The photon state}

We can now address the question of interest: What is the state of the photon emitted from the cavity to the detected mode? The photon state is computed in two steps. First, we obtain the sub-normalized wave-function of a photon in the pure state conditioned on the atom-cavity system being in the state $\ket{s,0}$ at time $s$ and no scattering events $L_2$ and $L_3$ happening at later times. Second, we solve the full master equation to compute the probability of a scattering via $L_2$ happening at time $s.$ Such a scattering event projects the system back onto $\ket{s,0}$.

\subsubsection{Conditional pure photon wave-function}
Let us start by addressing the wave-function of a photon in a pure state (pure photon) conditioned on the atom-cavity system being in the state $\ket{s,0}$ at time $s$ and no scattering events $L_2$ and $L_3$ happening at later times. To do so, we need to solve the evolution of the atom-cavity system conditioned to the case with no scattering. The Lindbladian part of the master equation~\eqref{eq:ME} describes random noise processes affecting the system. In particular, the terms $L_i \varrho_t L_i^\dag dt$ corresponds to a scattering happening during an infinitesimal time interval $dt$. In contrast, the conjugate term  $-\frac{1}{2}\{L_i^\dag L_i, \varrho_t\} dt$ corresponds to no scattering happening during $dt.$ Its role can be thought of as reducing the probability to find the system in the pre-scattered state $\varrho_{t+dt} \to \varrho_{ t} - \frac{1}{2}\{L_i^\dag L_i, \varrho_{t}\}dt$. Hence, to describe the evolution of the system conditioned on no scattering, we drop all the terms $L_i\varrho_t L_i^\dag$ but keep their conjugate terms $-\frac{1}{2}\{L_i^\dag L_i, \varrho_t\}$ in the master equantion~\eqref{eq:ME}. It is straightforward to see that such an evolution preserves the purity of a state, and can be written in the form of the Schr\"odinger equation with a non-Hermitian Hamiltonian
\be
\dot{\ket{ \Phi_t}} = \left(-\ii H_t - \frac{1}{2} \sum_i L_i^\dag L_i \right) \ket{\Phi_t}.
\ee
With the initial condition $\ket{ \Phi_s}=\ket{s,0},$ this equation can be solved to give the system state $\ket{\Phi_{t|s}}$ conditioned on the event corresponding to no scattering at time $t\geq s$. In particular, for a (piece-wise) constant Rabi frequency, the solution reads
$\ket{\Phi_{t|s}}=e^{\left(-\ii H - \frac{1}{2} \sum_i L_i^\dag L_i \right) (t-s)}\ket{s,0}$ which can be computed numerically. To obtain the amplitude of the emitted photon at a given time we project the atom-cavity state at this time into
$\sqrt{2\kappa}\bra{d,1}$.  In the laboratory frame, this gives
\be\label{eq:pure photon}
\begin{split}
	\ket{\psi_s} &= \int_{s}^{\infty}  \psi_s(t) \, a_t^\dag \ket{0} dt,\\
	\psi_s(t)&=  \sqrt{2 \kappa}\,  e^{-\ii \omega_C t}\, \braket{d,1}{\Phi_{t|s}}.
\end{split}
\ee
The photonic state $\ket{\psi_s}$ is sub-normalized. Its norm  
\be
p_{pure}(s)=\braket{\psi_s}{\psi_s}
\ee
is precisely the probability that no scattering event happens after time $s$ (given the initial condition). The conditional state thus reads $\ket{\psi_s}/\sqrt{p_{pure}(s)}$. One notes that $p_{pure}(0)$ is the probability that a photon is emitted without a single scattering during the evolution.

\subsubsection{Scattering probability}
We now solve the full master equation~\eqref{eq:ME} and obtain the atom-cavity state $\varrho_t$ for all times. Note that for a (piecewise) constant Rabi frequency, the solution can be obtained analytically by vectorizing the master equation and the density matrix. From this state, we compute the probability of scattering back to $\ket{s,0}$ at time $s$
\be\label{eq:scattering rate}
P(s) = \tr{\varrho_s \, L_2^\dag L_2}.
\ee

\subsubsection{The photon state}
The state of the emitted photon is given by
\be\label{eq: ph state1}
\rho = \prjct{\psi_0} + \int_{0}^\infty P(s) \prjct{\psi_s} ds + P_0 \prjct{0},
\ee
where  $\ket{\psi_s}$ is given in Eq.~\eqref{eq:pure photon} and $P(s)$ in Eq.~\eqref{eq:scattering rate}. Let us comment on each contribution separately. The first term $\prjct{\psi_0}$ describes a pure photon emitted without a single scattering ($L_2, L_3$). This happens with probability $p_{pure}(0)$. The integral collects all the possibilities where the last $p-s$-scattering happens at time $s$ and no scattering events happen at later times. Any such history happens with probability $P(s)p_{pure}(s)$ and yields a pure photon in the state $\ket{\psi_s}/\sqrt{p_{pure}(s)}$.  Finally, the last terms $P_0\prjct{0}$ with 
\be\begin{split}
	P_0&=1-\tr{\prjct{\psi_0} + \int_{0}^\infty P(s) \prjct{\psi_s} ds }\\
	& = 1- p_{pure}(0) -\int P(s)p_{pure}(s) ds 
\end{split}\ee 
collects the cases where no photon is emitted from the cavity. If the laser pulse is not turned off, this term can be alternatively computed as the overall probability of the $p-d$ scattering 
\be
P_0= \int_0^\infty \textrm{tr}(L_3^\dag L_3\,  \rho_t).
\ee 
To shorten the equations we will combine the first two contributions of Eq.~\eqref{eq: ph state1} together by defining
\be
\bar P(s)= P(s)+2 \delta(s),
\ee
where $\delta(s)$ is the delta-function with $2 \int_0^\epsilon \delta(s)ds = 1$. The photon state $\rho$ in Eq.\eqref{eq: ph state1} then simply reads
\be\label{eq:photon state}
\rho = \int_{0}^\infty \bar P(s) \prjct{\psi_s} ds + P_0 \prjct{0}.
\ee

\subsubsection{Expected number of scattering events}
Note that the average number of $L_2$ scattering events per experimental run is simply given by the time integral of the scattering rate
\be
\int_0^\infty P(s) ds,
\ee
and equals the expected number of laser photons scattered on the $p-s$ transition.
\subsection{Photon statistics}

Now that we have computed the photonic state $\rho$ emitted from the cavity, let us consider the Hong-Ou-Mandel (HOM) interference of two such photons, as depicted in Fig.~1 of the main text. More precisely, we consider two photons described by Eq~\eqref{eq:photon state} that enter the two ports of a 50/50 beamsplitter followed by two photon detectors D1 and D2. We assume detection with unit efficiency here, and discuss the general case in the next section.

\subsubsection{Single click rates}
The probability density that a photon in the state $\rho$ given in Eq.~\eqref{eq:photon state} triggers a click on the detector D1 (the same result is obtained for the detector D2) at time $t$ is given by $p_s(t)=\frac{1}{2}\tr{ \rho\,  a_{t}^\dag \prjct{0}a_t}$. Here, the $1/2$ factor comes from $50/50$ beamsplitter. Direct application of Eqs.~\eqref{eq:photon state} and \eqref{eq:pure photon} gives
\be
p_{s}(t) = \frac{1}{2} \int_{0}^\infty \bar P(s) |\psi_s(t)|^2 ds,
\ee
where we formally set $\psi_s(t) =0$ for $s>t$ here and in the following. 

\subsubsection{Coincidence rates}
We can now compute the twofold coincidence rate when one photon is sent at each input of the beamsplitter. For two photons with orthogonal polarizations corresponding to states $\rho_a$ and $\rho_{b_\bot}$, the probability to get a click at time $t_1$ in the detector D1 and a click at time $t_2$ in the detector D2 is given by
\be
p_{C\perp}(t_1,t_2) = p_S^{(a)}(t_1) p_S^{(b_\bot)}(t_2)+p_S^{(a)}(t_2) p_S^{(b_\bot)}(t_1)\\
\ee
simply because there is no interference. \\

Next, we consider two photons with the same polarization $\rho_a$ and $\rho_b.$ They are respectively characterized by $\bar P_a(s_a)$ with $\psi^{(a)}_{s_a}$ and $\bar P_b(s_b)$ with $\psi^{(b)}_{s_b}$ accordingly to Eq.~\eqref{eq:photon state}.  The twofold coincidence probability with the detector D1 clicking at time $t_1$ and the  detector D2 clicking at $t_2$ is computed as $p_c(t_1,t_2)=\tr{\rho_a\otimes\rho_b\,  \Pi_{t_1,t_2}}$ where $\Pi_{t_1,t_2}$ is the projector onto 
\be
\frac{1}{2}(a_{t_1}^\dag+b_{t_1}^\dag)(a_{t_2}^\dag - b_{t_2}^\dag)\ket{00}
\ee
where $a_{t_1}^\dag$ for example is the bosonic creation operator for the input mode $a$ at time $t_1$, as in Eq.\eqref{eq:pure photon}. From Eqs.~\eqref{eq:photon state} and \eqref{eq:pure photon} we find
\be\label{eq:pcparallel}\begin{split}
	p_{C\parallel}(t_1,t_2)&=
	\frac{1}{4}\int_0^\infty \bar P_a(s_a) \bar P_b(s_b)\times\\
	&|\psi_{s_a}^{(a)}(t_1)\psi_{s_b}^{(b)}(t_2)- \psi_{s_a}^{(a)}(t_2)\psi_{s_b}^{(b)}(t_1)|^2 ds_a ds_b.
\end{split}
\ee
The integrand in the last equation reads
\be\begin{split}\label{eq: bunching}
	&|\psi_{s_a}^{(a)}(t_1)\psi_{s_b}^{(b)}(t_2)- \psi_{s_a}^{(a)}(t_2)\psi_{s_b}^{(b)}(t_1)|^2 =\\
	&|\psi_{s_a}^{(a)}(t_1)|^2 |\psi_{s_b}^{(b)}(t_2)|^2 + |\psi_{s_a}^{(a)}(t_2)|^2 |\psi_{s_b}^{(b)}(t_1)|^2\\
	&- \psi_{s_a}^{(a)}(t_1)\psi_{s_a}^{(a)*}(t_2) \,\psi_{s_b}^{(b)*}(t_1)\psi_{s_b}^{(b)}(t_2)\\
	&- \psi_{s_a}^{(a)*}(t_1)\psi_{s_a}^{(a)}(t_2)\, \psi_{s_b}^{(b)}(t_1)\psi_{s_b}^{(b)*}(t_2).
\end{split}\ee
The last two terms are responsible for a destructive interference and bunching of the incident photons, which reduces the coincidence rate as compared to the orthogonal case.

\subsubsection{Visibility of the Hong-Ou-Mandel pattern}

The absolute detection times $t_1$ and $t_2$ are not relevant for computing the visibility of the Hong-Ou-Mandel (HOM) interference pattern. What matters is the difference between the detection times
\be
\tau = t_1-t_2.
\ee
We define the coincidence rate as a function of this delay between the clicks
\be
p_{C\parallel(\perp)}(\tau) = \int p_{C\parallel(\perp)}(t_2+\tau,t_2 )\, dt_2.\\
\ee
Furthermore, we define detection window $T$, by accepting only the coincidence events with a delay $|\tau|\leq T$ falling inside this window. For a fixed window $T$ the HOM visibitity is defined as 
\be
V(T) = 1-R(T)
\ee
where $R(T)$ is the ratio between the  coincidence probabilities for parallel and orthogonal polarizations with a bounded delay $|\tau|<T$
\be
R(T) =\frac{\int_{-T}^T p_{C\parallel}(\tau)  d\tau}{\int_{-T}^T p_{C\perp}(\tau)  d\tau }.\\
\ee

As mentioned after Eq.~\eqref{eq: bunching}, the effect of the photon bunching on the visibility is captured by 
\be\begin{split}
	\int_{-T}^T d\tau \int d{t_2} \, &\psi_{s_a}^{(a)}(t_2+\tau)\psi_{s_a}^{(a)*}(t_2) \,\psi_{s_b}^{(b)*}(t_2+\tau )\psi_{s_b}^{(b)}(t_2)\\
	&
	\nonumber
	+ h.c.
\end{split}
\ee
From this equation, we see that $p_{C\parallel}(0)=0.$ The visibility thus increases when the detection window $T$ is shortened. This can be intuitively understood from the fact that narrowing the detection window removes some of the temporal mixedness of the photons and effectively purifies them. The price to pay for decreasing $T$ is that the probability 
\be\label{eq:Psucc}
P_\text{succ}(T) =  \int_{-T}^T p_{C\perp}(\tau) d\tau.
\ee
to observe a coincidence within the detection window decreases with $T$ for orthogonally polarized photons.

\subsection{Predictions}

\begin{figure}
	\centering
	\includegraphics[width=0.4 \textwidth]{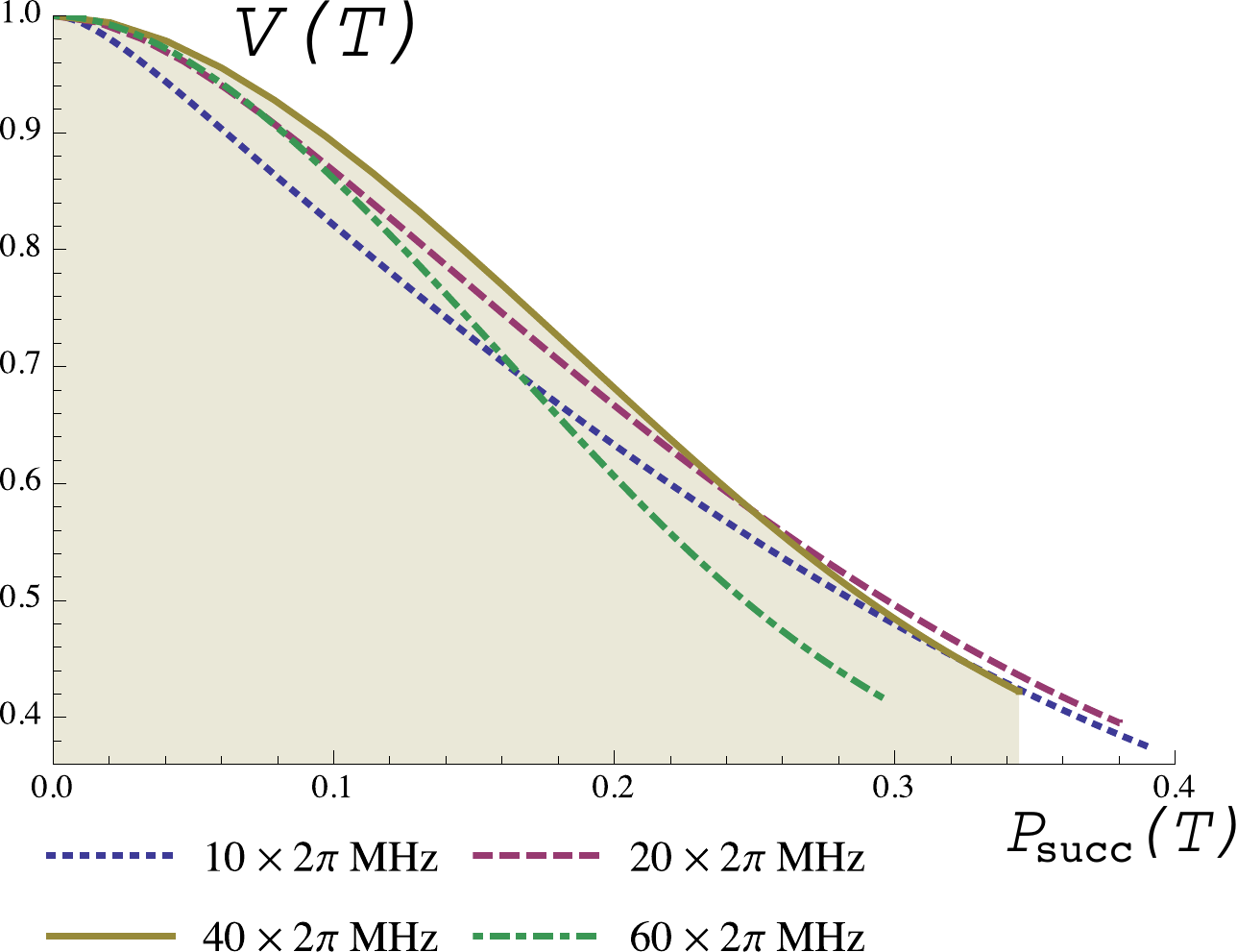}
	\caption{A parametric plot of the Hong-Ou-Mandel visibility $V(T)$ versus the success probability $P_\text{succ}(T)$ for different values of the Rabi frequency $\Omega$. For each curve $T\in[0,\infty)$ is increasing from left to right. 
		All the feagures are computed for $\Delta= - 2\pi \times 400$ MHz, $\kappa=2\pi \times 0.07$ MHz, $\gamma_{sp} = 2\pi \times 10.7$ MHz, $\gamma_{dp}= 2\pi \times 0.7$ MHz, and $g=2\pi \times 1.2 \sqrt{4/15}$ MHz.}
	\label{fig:VP}
\end{figure}

With such a theoretical model for the state of emitted photons and the HOM interference pattern, we can find experimental parameters facilitating the implementation of certain tasks that are relevant for quantum networking. An interesting experiment in this framework aims to entangle two ions remotely by first creating locally ion-photon entanglement and then performing a photonic Bell state measurement at a central station \cite{Matsukevich2008, Moehring2007}. Ion-photon entanglement is created by modifying our experiment to enable the transitions from the excited state $\ket{p}$ to two states $\ket{d}$ and $\ket{d'}$ leading to cavity photons with orthogonal polarizations, as we have previously shown \cite{Stute2012, 50km}. Two such photons emitted from cavities located at different locations are then sent into a beamsplitter which is followed by two detectors. A twofold coincidence projects the two ions into an entangled state. In such an entanglement swapping experiment, the fidelity $F$ (with respect to a maximally entangled two qubit state) of the two ion state can be shown to be proportional to the HOM visibility $V$ \cite{PhysRevLett.123.213601}. Furthermore, the rate at which the entangled states are created is related to the success probability $P_\text{succ}$ defined in Eq.~\eqref{eq:Psucc}. Hence, both the visibility $V$ and the success rate $P_\text{succ}$ play an important role for the entanglement swapping experiment. We thus wish to find experimental parameters maximizing the success rate given that the visibility stays above a threshold value (or vice versa). In particular, we focus on the impact of the Rabi frequency $\Omega$ on the visibility $V$ and the success rate $P_\text{succ}$. In Figure \ref{fig:VP}, we give a parametric plot of $V$ and $P_\text{succ}$ for various detection window $T$. Different curves correspond to different values of the Rabi frequency $\Omega_t=\Omega$ for $t\geq 0$. We see that in the range of small $P_\text{succ}$ where $V$ remains high, there is a globally optimal Rabi frequency $\Omega_\text{opt}\approx 2\pi \times 40$ MHz which maximizes $V$ for all values of $P_\text{succ}$. \\

\subsection{Extended models: imperfections beyond scattering}

In this section we discuss various imperfections that are detrimental for the photon counting statistics, and show how to describe them within our model.

\subsubsection{Detection efficiency}
The non-unit detection efficiency is modelled by a loss channel acting on the photon state prior to detection. Since the photonic state $\rho$ in Eq.~\eqref{eq:photon state} is a mixture of a single photon distributed across difference temporal modes and a vacuum component, the effect of loss is particularly simple to describe. A loss channel with transmission rate $\eta$ simply maps
\be
\bar P(s) \mapsto \eta \bar P(s)\quad P_0 \mapsto 1-\eta(1-P_0).
\ee
The loss of cavity photons outside of the detected mode is also equivalent to a lack of detection efficiency. In such a case, the total cavity decay rate entering in the master equation is the sum of the loss rate towards the detector $\kappa_\text{det}$ and in modes that are not measured $\kappa_\text{loss}$, that is 
$\kappa = \kappa_\text{det} + \kappa_\text{loss}.$
The detection efficiency is reduced by an additional factor $\eta_\kappa=\frac {\kappa_\text{det}} \kappa$.

\subsubsection{Mode mismatch}
In practice, the two modes entering the beamsplitter preceding the two detectors do not perfectly overlap, which reduces the visibility of their interference pattern. An imperfect overlap of $\varepsilon$ means that with
probability $\varepsilon$ the two photons entering the interference will not see each other. The coincidence rate for two photons with the same polarization becomes
\be
\label{dist}
p_{C\parallel}\to (1-\varepsilon)\, p_{C\parallel} + \varepsilon\, p_{C\perp}.\\
\ee
This is how we model a fixed nonzero distinguishability of the interfering photons in order to reproduce experimental date, see Figs.~2 and 3 in the main text. 

\subsubsection{Photon frequency mismatch}
We here model the effect of the frequency distinguishability of the two photons arriving at the beamsplitter on the HOM-type interference experiment. We restrict ourselves to a simple approximation where the photons have a constant offset and a linear drift of the frequency in time    
\be
\label{eq_carrier_offset}
\omega(t) = \omega_0 + \omega_s + \omega_d t
\ee
where $\omega_s$ and $\omega_d$ can be random variables over the experimental attempts. We will come to their distributions later.\\

In the two photon experiment, the frequency uncertainty affects the phase of the photonic wave-function. In particular, for the computation of the coincidence rate, we are interested in the factor 
\be
f=\mean{e^{\ii (\phi_1(t_1)-\phi_1(t_2)-\phi_2(t_1)+\phi_2(t_2))}},
\ee
where $\phi_1$ and $\phi_2$ is the phase of photons emitted at times $t_1$ and $t_2,$ and $\mean{\bullet}$ denotes the statistical average over the phases. The photon phase is the time integral of its instantaneous frequency
\be
\phi(t) = \int_0^t \omega(s)ds = \omega_s t + \omega_d \frac{t^2}{2}.
\ee
Given that the second photon is delayed from the first one by a time $\tau,$ we get
\be\begin{split}
	&(\phi_1(t_1)-\phi_1(t_2)-\phi_2(t_1)+\phi_2(t_2)) \\
	&= (\phi(t_1)-\phi(t_2)-\phi(t_1+\tau)+\phi(t_2+\tau)) \\
	& = \omega_d \tau (t_2-t_1).
\end{split}\ee
and thus 
\be\begin{split}
	f &= \int e^{\ii  \omega_d \tau (t_2-t_1)}p(\omega_d) d\omega_d\\
	&=  \tilde p(\tau (t_2-t_1)),
\end{split}
\ee
where $\tilde p$ is the Fourier transform of $p(\omega_d)$. Let us assume that the distribution of the drift is a Gaussian with a zero mean and an unknown variance $\sigma^2$ 
\be
p(\omega_d) = \frac{1}{\sqrt{2\pi} \sigma} e^{-\frac{\omega_d^2}{2 \sigma^2}}.
\ee
In this case, 
\be
f= e^{-\frac{1}{2}\tau^2(t_2-t_1)^2 \sigma^2}.
\ee
It remains to relate $\sigma$ to experimentally measured parameters. In the experiment, we measure $v(\bar T)$ -- the average squared deviation of the cavity frequency over a time $\bar T$. Under our assumption, this quantity reads
\be
\label{freq_drift}
v(\bar T) = \mean{\big(\omega(0) -\omega(\bar T)\big)^2} = \bar{T}^2 \sigma^2
.\ee
We obtain
\be
\begin{split}
	f(|t_2-t_1|,\tau) &= e^{-\frac{1}{2}\tau^2(t_2-t_1)^2\frac{v(\bar T)}{\bar{T}^2}}\\
	&\approx 1-\frac{ \tau^2(t_2-t_1)^2 v(\bar{T})}{2 \bar{T}^2}.
\end{split}
\ee
Finally, let us note that we do not need to assume the exact form of the distribution $p(\omega_d)$. As long as the distribution is symmetric $p(\omega_d)=p(-\omega_d)$, and the dephasing effect is small, we can develop the Fourier transform of the distribution $p(\omega_d)$ to the second order to get 
\be
\tilde{p}(\tau (t_2-t_1))\approx  1- \tau^2(t_2-t_1)^2 \sigma^2
\ee
as a function of a single parameter $\sigma^2$, which is expressed in the same way as $v(\bar{T})/\bar{T}^2$.

\section{Data modelling}

\subsection{General principles:} 

\subsubsection{Spontaneous scattering rates}

For all modelling, the scattering rates are  $\gamma_{pd}=2\pi\times0.68$~MHz and $\gamma_{ps}=2\pi\times10.7$~MHz  (Figure 1, main text). 

\subsubsection{g factors, Clebsch-Gordan factors and finite temperature effects}

Following \cite{russothesis}, by adiabatically 
eliminating the $P$-state population, the coherent part of the cavity-mediated Raman transition (CMRT) can be described as an effective two-level 
system driven with $\Omega_{eff} = \frac{\Omega^{\prime}\beta g_0}{\Delta+\delta}$.
Here, $\Omega^\prime$ is the drive strength of the CMRT (defined later),  $\beta$ is the product of the Clebsch-Gordan-coefficient of the 854~nm atomic transition and the projection of the polarization plane of the cavity mode onto
the atomic dipole moment, $\Delta+\delta$ is the detuning (\rfig{fig_levelscheme}) and $g_0$ is the maximum ion-cavity 
coupling strength. 
The finite temperature of the ion after Doppler cooling leads to the Raman laser coupling to the ion motional sidebands and 
thus a reduced coupling strength on the desired carrier transition $\Omega^\prime = \alpha\Omega$. Here, $\alpha$  (a real number, less than 1) is the reduction factor of the carrier drive-strength and $\Omega$ is the Raman-laser Rabi frequency used for the theory calculations (eq. \ref{eq:Hamil}) and throughout the manuscript. 
We calculate  $\alpha$ using eq. 3.11 of \cite{roosthesis} for the known (independently measured) motional state of the ion in all directions.
We define the effective ion-cavity coupling $g = \alpha \beta g_0$, which is used for the theoretical calculations (eq. \ref{eq:Hamil}).
For the transition to the $\mathrm{D_{m_j=-5/2}}$ state (V photon, long path) $\beta = \sqrt{10/15\cdot 1/2}$ , where 1/2 stands for the projection of the polarization plane of the cavity mode to the atomic dipole moment. For the transition to the $\mathrm{D_{m_j=-3/2}}$ state (H photon, short path) $\beta = \sqrt{4/15}$.

\subsubsection{Calibration of experimental parameters}

The maximum ion-cavity coupling strength $g_0$ is calculated using the known cavity length and waist, and indirectly measured by comparing photon wave-packet and generation efficiency (measured at negligible temperature, with additional side-band cooling) to numerical 
simulations \cite{russothesis}. Both approaches give a value of $g_0/2\pi = 1.53\pm0.01~$ MHz.
We calculate that the ion's wavepacket delocalisation after Doppler cooling introduces no significant (compared to $\alpha$) decrease of the ion-cavity coupling strength (the delocalisation is predominantly in a direction perpendicular to the vacuum cavity standing waves). 
We determine $\Omega$ by performing 393~nm spectroscopy of the Raman resonance for 
different intensities of the Raman beam and extracting the induced AC-Stark shift.
The cavity detuning ($\Delta=403\pm 5$ MHz), measured with a wavemeter, is the frequency difference of the 393 laser when 
tuned on resonance with the S-P transition and when tuned to the Raman 
resonance condition. The lifetime of the P state limits the precision of this measurement.

We determine the mean phonon number of each motional modes of the ion by performing Rabi 
flops on the 729 ($S_{1/2}\rightarrow D_{5/2}$) transition and fitting the observed dependence of the
excitation probability on the pulse length with a model that takes into account the ion temperature (for details see \cite{roosthesis}).  
Flops are taken with two different 729 nm laser beam directions, allowing the temperature in different motional modes to be distinguished.  

\subsection{Modelling in figure 2} 

\subsubsection{Ideal model}

For the theory curves presented in Figure 2 of the main text, 
we experimentally determine $\Omega/2\pi = 63.5(5)$ MHz, $\Delta/2\pi = 403(5)~$MHz. 
The coupling-reducing factor $\alpha = 0.75(2)$ ($g/(2\pi \beta) = 1.15~$MHz) was calculated from a mean-phonon-number of 14 on the axial mode ($\omega_{ax} = 0.9$ MHz) and a mean-phonon-number of 6 on the radial mode ($\omega_{rad} = 2.4$ MHz).

\subsubsection{Extended model}

For the extended-model curves in Figure 2 we model the frequency mismatch as a constant frequency offset according to eq. \ref{eq_carrier_offset}, with $\omega_s/2\pi=40~$ kHz, $\omega_d = 0$ (from best match to the experimental data). 

Due to the excess of distinguishability in the experiment compared to theory
even for the smallest coincidence window $T$, an additional constant 
distinguishability of $\epsilon=0.01$ was introduced into the model, according to 
\ref{dist}. This could be caused by any mode mismatch, e.g. imperfect 
photon polarizations at the final beam-splitter due to an imperfectly aligned PBS and slight imperfections in the beam splitter itself. The effect of background coincidences (due to dark 
counts, background light, imperfect $g^{(2)}(0)$) was measured to be at the 
$5\times 10^{-4}$ level and was neither subtracted from the experiment data nor 
taken into account in the model.  

\subsection{Modelling in figure 3: frequency converted case} 

\subsubsection{Ideal model}
For the theory curves presented in Figure 3 (main text) we
determine $\Omega/2\pi = 64.3(5)$ MHz, $\Delta/2\pi = 403(5)~$MHz. The coupling reducing factor $\alpha = 0.69(2)$ ($g/(2\pi\beta) = 1.05~$MHz) was determined from the 
best fit of the simulated photon wavepacket. \
This value is lower than the value expected from the temperature calibration measurement before the experiment ($g/(2\pi\beta) = 1.15~$MHz) which could be caused by slight drift of the Doppler laser cooling parameters. 

\subsubsection{Extended model}
As in the case of the experiment without conversion, the discrepancy of theoretical and experimental visibilities, which depends on the coincidence window size, is attributed to the photons' frequency mismatch. 
When implementing photon frequency conversion before and after the delay line, 
the effect of the frequency instability of the unstablized conversion pump laser (1902 nm) 
has to be taken into account and is the dominant effect. 
In our 
implementation there is a significant optical path length difference of 
$\sim15 \mu$s (the whole delay line) between the pump laser field and the photon 
(see \rfig{fig:setup_full}). In that case any laser frequency drift on this time 
scale will result in a frequency mismatch of the short and long path photons 
at the interference beamsplitter. Note, however, that the constant (in time) part of the pump-laser frequency is expected to cancel out since the down and up conversion processes are symmetric. Based on the laser specification we expect $\sim 50$ kHz instability on a $10 \mu$s timescale, making it the dominant source of frequency mismatch.

To plot the extended-model curves in Figure 3 (main text) we use 
\ref{freq_drift} with $v(10\mu s) = 2\pi\cdot 50~$kHz and take into account the measured background-coincidence-rate by adding a constant coincidence-probability-density of $0.8\cdot 10^{-6}~\mu s^{-1}$ (Figure 3b). 
The extended theory lines of Figure 3c are calculated by integrating the 
coincidence distributions including this background floor. The background 
coincidences are significant in the experiment with conversion because 
the signal level is $~5$ times lower. Also, the observed free-running noise counts were 13(2) cps compared to 2(2) cps per detector for the experiment 
without conversion due to the photon noise introduced by the frequency 
conversion process.


\begin{thebibliography}{43}%
\makeatletter
\providecommand \@ifxundefined [1]{%
 \@ifx{#1\undefined}
}%
\providecommand \@ifnum [1]{%
 \ifnum #1\expandafter \@firstoftwo
 \else \expandafter \@secondoftwo
 \fi
}%
\providecommand \@ifx [1]{%
 \ifx #1\expandafter \@firstoftwo
 \else \expandafter \@secondoftwo
 \fi
}%
\providecommand \natexlab [1]{#1}%
\providecommand \enquote  [1]{``#1''}%
\providecommand \bibnamefont  [1]{#1}%
\providecommand \bibfnamefont [1]{#1}%
\providecommand \citenamefont [1]{#1}%
\providecommand \href@noop [0]{\@secondoftwo}%
\providecommand \href [0]{\begingroup \@sanitize@url \@href}%
\providecommand \@href[1]{\@@startlink{#1}\@@href}%
\providecommand \@@href[1]{\endgroup#1\@@endlink}%
\providecommand \@sanitize@url [0]{\catcode `\\12\catcode `\$12\catcode
  `\&12\catcode `\#12\catcode `\^12\catcode `\_12\catcode `\%12\relax}%
\providecommand \@@startlink[1]{}%
\providecommand \@@endlink[0]{}%
\providecommand \url  [0]{\begingroup\@sanitize@url \@url }%
\providecommand \@url [1]{\endgroup\@href {#1}{\urlprefix }}%
\providecommand \urlprefix  [0]{URL }%
\providecommand \Eprint [0]{\href }%
\providecommand \doibase [0]{http://dx.doi.org/}%
\providecommand \selectlanguage [0]{\@gobble}%
\providecommand \bibinfo  [0]{\@secondoftwo}%
\providecommand \bibfield  [0]{\@secondoftwo}%
\providecommand \translation [1]{[#1]}%
\providecommand \BibitemOpen [0]{}%
\providecommand \bibitemStop [0]{}%
\providecommand \bibitemNoStop [0]{.\EOS\space}%
\providecommand \EOS [0]{\spacefactor3000\relax}%
\providecommand \BibitemShut  [1]{\csname bibitem#1\endcsname}%
\let\auto@bib@innerbib\@empty
\bibitem [{\citenamefont {Kimble}(2008)}]{Kimble2008}%
  \BibitemOpen
  \bibfield  {author} {\bibinfo {author} {\bibfnamefont {H.~J.}\ \bibnamefont
  {Kimble}},\ }\href {https://doi.org/10.1038/nature07127} {\bibfield
  {journal} {\bibinfo  {journal} {Nature}\ }\textbf {\bibinfo {volume} {453}},\
  \bibinfo {pages} {1023} (\bibinfo {year} {2008})}\BibitemShut {NoStop}%
\bibitem [{\citenamefont {Wehner}\ \emph {et~al.}(2018)\citenamefont {Wehner},
  \citenamefont {Elkouss},\ and\ \citenamefont {Hanson}}]{Wehnereaam9288}%
  \BibitemOpen
  \bibfield  {author} {\bibinfo {author} {\bibfnamefont {S.}~\bibnamefont
  {Wehner}}, \bibinfo {author} {\bibfnamefont {D.}~\bibnamefont {Elkouss}}, \
  and\ \bibinfo {author} {\bibfnamefont {R.}~\bibnamefont {Hanson}},\ }\href
  {http://science.sciencemag.org/content/362/6412/eaam9288} {\bibfield
  {journal} {\bibinfo  {journal} {Science}\ }\textbf {\bibinfo {volume} {362}}
  (\bibinfo {year} {2018})}\BibitemShut {NoStop}%
\bibitem [{\citenamefont {Duan}\ \emph {et~al.}(2001)\citenamefont {Duan},
  \citenamefont {Lukin}, \citenamefont {Cirac},\ and\ \citenamefont
  {Zoller}}]{Duan2001}%
  \BibitemOpen
  \bibfield  {author} {\bibinfo {author} {\bibfnamefont {L.-M.}\ \bibnamefont
  {Duan}}, \bibinfo {author} {\bibfnamefont {M.~D.}\ \bibnamefont {Lukin}},
  \bibinfo {author} {\bibfnamefont {J.~I.}\ \bibnamefont {Cirac}}, \ and\
  \bibinfo {author} {\bibfnamefont {P.}~\bibnamefont {Zoller}},\ }\href
  {https://doi.org/10.1038/35106500} {\bibfield  {journal} {\bibinfo  {journal}
  {Nature}\ }\textbf {\bibinfo {volume} {414}},\ \bibinfo {pages} {413}
  (\bibinfo {year} {2001})}\BibitemShut {NoStop}%
\bibitem [{\citenamefont {Monroe}\ \emph {et~al.}(2014)\citenamefont {Monroe},
  \citenamefont {Raussendorf}, \citenamefont {Ruthven}, \citenamefont {Brown},
  \citenamefont {Maunz}, \citenamefont {Duan},\ and\ \citenamefont
  {Kim}}]{PhysRevA.89.022317}%
  \BibitemOpen
  \bibfield  {author} {\bibinfo {author} {\bibfnamefont {C.}~\bibnamefont
  {Monroe}}, \bibinfo {author} {\bibfnamefont {R.}~\bibnamefont {Raussendorf}},
  \bibinfo {author} {\bibfnamefont {A.}~\bibnamefont {Ruthven}}, \bibinfo
  {author} {\bibfnamefont {K.~R.}\ \bibnamefont {Brown}}, \bibinfo {author}
  {\bibfnamefont {P.}~\bibnamefont {Maunz}}, \bibinfo {author} {\bibfnamefont
  {L.-M.}\ \bibnamefont {Duan}}, \ and\ \bibinfo {author} {\bibfnamefont
  {J.}~\bibnamefont {Kim}},\ }\href {\doibase 10.1103/PhysRevA.89.022317}
  {\bibfield  {journal} {\bibinfo  {journal} {Phys. Rev. A}\ }\textbf {\bibinfo
  {volume} {89}},\ \bibinfo {pages} {022317} (\bibinfo {year}
  {2014})}\BibitemShut {NoStop}%
\bibitem [{\citenamefont {K{\'o}m{\'a}r}\ \emph {et~al.}(2014)\citenamefont
  {K{\'o}m{\'a}r}, \citenamefont {Kessler}, \citenamefont {Bishof},
  \citenamefont {Jiang}, \citenamefont {S{\o}rensen}, \citenamefont {Ye},\ and\
  \citenamefont {Lukin}}]{Komar2014}%
  \BibitemOpen
  \bibfield  {author} {\bibinfo {author} {\bibfnamefont {P.}~\bibnamefont
  {K{\'o}m{\'a}r}}, \bibinfo {author} {\bibfnamefont {E.~M.}\ \bibnamefont
  {Kessler}}, \bibinfo {author} {\bibfnamefont {M.}~\bibnamefont {Bishof}},
  \bibinfo {author} {\bibfnamefont {L.}~\bibnamefont {Jiang}}, \bibinfo
  {author} {\bibfnamefont {A.~S.}\ \bibnamefont {S{\o}rensen}}, \bibinfo
  {author} {\bibfnamefont {J.}~\bibnamefont {Ye}}, \ and\ \bibinfo {author}
  {\bibfnamefont {M.~D.}\ \bibnamefont {Lukin}},\ }\href
  {https://doi.org/10.1038/nphys3000} {\bibfield  {journal} {\bibinfo
  {journal} {Nature Physics}\ }\textbf {\bibinfo {volume} {10}},\ \bibinfo
  {pages} {582} (\bibinfo {year} {2014})}\BibitemShut {NoStop}%
\bibitem [{\citenamefont {Khabiboulline}\ \emph {et~al.}(2019)\citenamefont
  {Khabiboulline}, \citenamefont {Borregaard}, \citenamefont {De~Greve},\ and\
  \citenamefont {Lukin}}]{PhysRevLett.123.070504}%
  \BibitemOpen
  \bibfield  {author} {\bibinfo {author} {\bibfnamefont {E.~T.}\ \bibnamefont
  {Khabiboulline}}, \bibinfo {author} {\bibfnamefont {J.}~\bibnamefont
  {Borregaard}}, \bibinfo {author} {\bibfnamefont {K.}~\bibnamefont
  {De~Greve}}, \ and\ \bibinfo {author} {\bibfnamefont {M.~D.}\ \bibnamefont
  {Lukin}},\ }\href {\doibase 10.1103/PhysRevLett.123.070504} {\bibfield
  {journal} {\bibinfo  {journal} {Phys. Rev. Lett.}\ }\textbf {\bibinfo
  {volume} {123}},\ \bibinfo {pages} {070504} (\bibinfo {year}
  {2019})}\BibitemShut {NoStop}%
\bibitem [{\citenamefont {Sekatski}\ \emph {et~al.}(2019)\citenamefont
  {Sekatski}, \citenamefont {W{\"o}lk},\ and\ \citenamefont
  {D{\"u}r}}]{sekatski2019optimal}%
  \BibitemOpen
  \bibfield  {author} {\bibinfo {author} {\bibfnamefont {P.}~\bibnamefont
  {Sekatski}}, \bibinfo {author} {\bibfnamefont {S.}~\bibnamefont {W{\"o}lk}},
  \ and\ \bibinfo {author} {\bibfnamefont {W.}~\bibnamefont {D{\"u}r}},\
  }\href@noop {} {\enquote {\bibinfo {title} {Optimal distributed sensing in
  noisy environments},}\ } (\bibinfo {year} {2019}),\ \Eprint
  {http://arxiv.org/abs/1905.06765} {arXiv:1905.06765 [quant-ph]} \BibitemShut
  {NoStop}%
\bibitem [{\citenamefont {Duan}\ and\ \citenamefont {Monroe}(2010)}]{Duan2010}%
  \BibitemOpen
  \bibfield  {author} {\bibinfo {author} {\bibfnamefont {L.-M.}\ \bibnamefont
  {Duan}}\ and\ \bibinfo {author} {\bibfnamefont {C.}~\bibnamefont {Monroe}},\
  }\href {\doibase 10.1103/RevModPhys.82.1209} {\bibfield  {journal} {\bibinfo
  {journal} {Rev. Mod. Phys.}\ }\textbf {\bibinfo {volume} {82}},\ \bibinfo
  {pages} {1209} (\bibinfo {year} {2010})}\BibitemShut {NoStop}%
\bibitem [{\citenamefont {Sangouard}\ \emph {et~al.}(2009)\citenamefont
  {Sangouard}, \citenamefont {Dubessy},\ and\ \citenamefont
  {Simon}}]{Sangouard2009}%
  \BibitemOpen
  \bibfield  {author} {\bibinfo {author} {\bibfnamefont {N.}~\bibnamefont
  {Sangouard}}, \bibinfo {author} {\bibfnamefont {R.}~\bibnamefont {Dubessy}},
  \ and\ \bibinfo {author} {\bibfnamefont {C.}~\bibnamefont {Simon}},\ }\href
  {\doibase 10.1103/PhysRevA.79.042340} {\bibfield  {journal} {\bibinfo
  {journal} {Phys. Rev. A}\ }\textbf {\bibinfo {volume} {79}},\ \bibinfo
  {pages} {042340} (\bibinfo {year} {2009})}\BibitemShut {NoStop}%
\bibitem [{\citenamefont {Reiserer}\ and\ \citenamefont
  {Rempe}(2015)}]{RevModPhys.87.1379}%
  \BibitemOpen
  \bibfield  {author} {\bibinfo {author} {\bibfnamefont {A.}~\bibnamefont
  {Reiserer}}\ and\ \bibinfo {author} {\bibfnamefont {G.}~\bibnamefont
  {Rempe}},\ }\href {\doibase 10.1103/RevModPhys.87.1379} {\bibfield  {journal}
  {\bibinfo  {journal} {Rev. Mod. Phys.}\ }\textbf {\bibinfo {volume} {87}},\
  \bibinfo {pages} {1379} (\bibinfo {year} {2015})}\BibitemShut {NoStop}%
\bibitem [{\citenamefont {Bruzewicz}\ \emph {et~al.}(2019)\citenamefont
  {Bruzewicz}, \citenamefont {Chiaverini}, \citenamefont {McConnell},\ and\
  \citenamefont {Sage}}]{reviewionsqc}%
  \BibitemOpen
  \bibfield  {author} {\bibinfo {author} {\bibfnamefont {C.~D.}\ \bibnamefont
  {Bruzewicz}}, \bibinfo {author} {\bibfnamefont {J.}~\bibnamefont
  {Chiaverini}}, \bibinfo {author} {\bibfnamefont {R.}~\bibnamefont
  {McConnell}}, \ and\ \bibinfo {author} {\bibfnamefont {J.~M.}\ \bibnamefont
  {Sage}},\ }\href {\doibase 10.1063/1.5088164} {\bibfield  {journal} {\bibinfo
   {journal} {Applied Physics Reviews}\ }\textbf {\bibinfo {volume} {6}},\
  \bibinfo {pages} {021314} (\bibinfo {year} {2019})}\BibitemShut {NoStop}%
\bibitem [{\citenamefont {Friis}\ \emph {et~al.}(2018)\citenamefont {Friis},
  \citenamefont {Marty}, \citenamefont {Maier}, \citenamefont {Hempel},
  \citenamefont {Holz\"apfel}, \citenamefont {Jurcevic}, \citenamefont
  {Plenio}, \citenamefont {Huber}, \citenamefont {Roos}, \citenamefont
  {Blatt},\ and\ \citenamefont {Lanyon}}]{Friis2018}%
  \BibitemOpen
  \bibfield  {author} {\bibinfo {author} {\bibfnamefont {N.}~\bibnamefont
  {Friis}}, \bibinfo {author} {\bibfnamefont {O.}~\bibnamefont {Marty}},
  \bibinfo {author} {\bibfnamefont {C.}~\bibnamefont {Maier}}, \bibinfo
  {author} {\bibfnamefont {C.}~\bibnamefont {Hempel}}, \bibinfo {author}
  {\bibfnamefont {M.}~\bibnamefont {Holz\"apfel}}, \bibinfo {author}
  {\bibfnamefont {P.}~\bibnamefont {Jurcevic}}, \bibinfo {author}
  {\bibfnamefont {M.~B.}\ \bibnamefont {Plenio}}, \bibinfo {author}
  {\bibfnamefont {M.}~\bibnamefont {Huber}}, \bibinfo {author} {\bibfnamefont
  {C.}~\bibnamefont {Roos}}, \bibinfo {author} {\bibfnamefont {R.}~\bibnamefont
  {Blatt}}, \ and\ \bibinfo {author} {\bibfnamefont {B.}~\bibnamefont
  {Lanyon}},\ }\href {\doibase 10.1103/PhysRevX.8.021012} {\bibfield  {journal}
  {\bibinfo  {journal} {Phys. Rev. X}\ }\textbf {\bibinfo {volume} {8}},\
  \bibinfo {pages} {021012} (\bibinfo {year} {2018})}\BibitemShut {NoStop}%
\bibitem [{\citenamefont {Brewer}\ \emph {et~al.}(2019)\citenamefont {Brewer},
  \citenamefont {Chen}, \citenamefont {Hankin}, \citenamefont {Clements},
  \citenamefont {Chou}, \citenamefont {Wineland}, \citenamefont {Hume},\ and\
  \citenamefont {Leibrandt}}]{PhysRevLett.123.033201}%
  \BibitemOpen
  \bibfield  {author} {\bibinfo {author} {\bibfnamefont {S.~M.}\ \bibnamefont
  {Brewer}}, \bibinfo {author} {\bibfnamefont {J.-S.}\ \bibnamefont {Chen}},
  \bibinfo {author} {\bibfnamefont {A.~M.}\ \bibnamefont {Hankin}}, \bibinfo
  {author} {\bibfnamefont {E.~R.}\ \bibnamefont {Clements}}, \bibinfo {author}
  {\bibfnamefont {C.~W.}\ \bibnamefont {Chou}}, \bibinfo {author}
  {\bibfnamefont {D.~J.}\ \bibnamefont {Wineland}}, \bibinfo {author}
  {\bibfnamefont {D.~B.}\ \bibnamefont {Hume}}, \ and\ \bibinfo {author}
  {\bibfnamefont {D.~R.}\ \bibnamefont {Leibrandt}},\ }\href {\doibase
  10.1103/PhysRevLett.123.033201} {\bibfield  {journal} {\bibinfo  {journal}
  {Phys. Rev. Lett.}\ }\textbf {\bibinfo {volume} {123}},\ \bibinfo {pages}
  {033201} (\bibinfo {year} {2019})}\BibitemShut {NoStop}%
\bibitem [{\citenamefont {Blinov}\ \emph {et~al.}(2004)\citenamefont {Blinov},
  \citenamefont {Moehring}, \citenamefont {Duan},\ and\ \citenamefont
  {Monroe}}]{Blinov2004}%
  \BibitemOpen
  \bibfield  {author} {\bibinfo {author} {\bibfnamefont {B.~B.}\ \bibnamefont
  {Blinov}}, \bibinfo {author} {\bibfnamefont {D.~L.}\ \bibnamefont
  {Moehring}}, \bibinfo {author} {\bibfnamefont {L.-M.}\ \bibnamefont {Duan}},
  \ and\ \bibinfo {author} {\bibfnamefont {C.}~\bibnamefont {Monroe}},\ }\href
  {https://doi.org/10.1038/nature02377} {\bibfield  {journal} {\bibinfo
  {journal} {Nature}\ }\textbf {\bibinfo {volume} {428}},\ \bibinfo {pages}
  {153} (\bibinfo {year} {2004})}\BibitemShut {NoStop}%
\bibitem [{\citenamefont {Moehring}\ \emph {et~al.}(2007)\citenamefont
  {Moehring}, \citenamefont {Maunz}, \citenamefont {Olmschenk}, \citenamefont
  {Younge}, \citenamefont {Matsukevich}, \citenamefont {Duan},\ and\
  \citenamefont {Monroe}}]{Moehring2007}%
  \BibitemOpen
  \bibfield  {author} {\bibinfo {author} {\bibfnamefont {D.~L.}\ \bibnamefont
  {Moehring}}, \bibinfo {author} {\bibfnamefont {P.}~\bibnamefont {Maunz}},
  \bibinfo {author} {\bibfnamefont {S.}~\bibnamefont {Olmschenk}}, \bibinfo
  {author} {\bibfnamefont {K.~C.}\ \bibnamefont {Younge}}, \bibinfo {author}
  {\bibfnamefont {D.~N.}\ \bibnamefont {Matsukevich}}, \bibinfo {author}
  {\bibfnamefont {L.-M.}\ \bibnamefont {Duan}}, \ and\ \bibinfo {author}
  {\bibfnamefont {C.}~\bibnamefont {Monroe}},\ }\href
  {https://doi.org/10.1038/nature06118} {\bibfield  {journal} {\bibinfo
  {journal} {Nature}\ }\textbf {\bibinfo {volume} {449}},\ \bibinfo {pages}
  {68} (\bibinfo {year} {2007})}\BibitemShut {NoStop}%
\bibitem [{\citenamefont {Hucul}\ \emph {et~al.}(2015)\citenamefont {Hucul},
  \citenamefont {Inlek}, \citenamefont {Vittorini}, \citenamefont {Crocker},
  \citenamefont {Debnath}, \citenamefont {Clark},\ and\ \citenamefont
  {Monroe}}]{Hucul:2015wo}%
  \BibitemOpen
  \bibfield  {author} {\bibinfo {author} {\bibfnamefont {D.}~\bibnamefont
  {Hucul}}, \bibinfo {author} {\bibfnamefont {I.~V.}\ \bibnamefont {Inlek}},
  \bibinfo {author} {\bibfnamefont {G.}~\bibnamefont {Vittorini}}, \bibinfo
  {author} {\bibfnamefont {C.}~\bibnamefont {Crocker}}, \bibinfo {author}
  {\bibfnamefont {S.}~\bibnamefont {Debnath}}, \bibinfo {author} {\bibfnamefont
  {S.~M.}\ \bibnamefont {Clark}}, \ and\ \bibinfo {author} {\bibfnamefont
  {C.}~\bibnamefont {Monroe}},\ }\href {\doibase 10.1038/nphys3150} {\bibfield
  {journal} {\bibinfo  {journal} {Nature Physics}\ }\textbf {\bibinfo {volume}
  {11}},\ \bibinfo {pages} {37} (\bibinfo {year} {2015})}\BibitemShut {NoStop}%
\bibitem [{\citenamefont {Stephenson}\ \emph {et~al.}(2019)\citenamefont
  {Stephenson}, \citenamefont {Nadlinger}, \citenamefont {Nichol},
  \citenamefont {An}, \citenamefont {Drmota}, \citenamefont {Ballance},
  \citenamefont {Thirumalai}, \citenamefont {Goodwin}, \citenamefont {Lucas},\
  and\ \citenamefont {Ballance}}]{balance}%
  \BibitemOpen
  \bibfield  {author} {\bibinfo {author} {\bibfnamefont {L.~J.}\ \bibnamefont
  {Stephenson}}, \bibinfo {author} {\bibfnamefont {D.~P.}\ \bibnamefont
  {Nadlinger}}, \bibinfo {author} {\bibfnamefont {B.~C.}\ \bibnamefont
  {Nichol}}, \bibinfo {author} {\bibfnamefont {S.}~\bibnamefont {An}}, \bibinfo
  {author} {\bibfnamefont {P.}~\bibnamefont {Drmota}}, \bibinfo {author}
  {\bibfnamefont {T.~G.}\ \bibnamefont {Ballance}}, \bibinfo {author}
  {\bibfnamefont {K.}~\bibnamefont {Thirumalai}}, \bibinfo {author}
  {\bibfnamefont {J.~F.}\ \bibnamefont {Goodwin}}, \bibinfo {author}
  {\bibfnamefont {D.~M.}\ \bibnamefont {Lucas}}, \ and\ \bibinfo {author}
  {\bibfnamefont {C.~J.}\ \bibnamefont {Ballance}},\ }\href@noop {} {\bibfield
  {journal} {\bibinfo  {journal} {arXiv preprint arXiv:1911.10841}\ } (\bibinfo
  {year} {2019})}\BibitemShut {NoStop}%
\bibitem [{\citenamefont {Stute}\ \emph {et~al.}(2012)\citenamefont {Stute},
  \citenamefont {Casabone}, \citenamefont {Schindler}, \citenamefont {Monz},
  \citenamefont {Schmidt}, \citenamefont {Brandst{\"a}tter}, \citenamefont
  {Northup},\ and\ \citenamefont {Blatt}}]{Stute2012}%
  \BibitemOpen
  \bibfield  {author} {\bibinfo {author} {\bibfnamefont {A.}~\bibnamefont
  {Stute}}, \bibinfo {author} {\bibfnamefont {B.}~\bibnamefont {Casabone}},
  \bibinfo {author} {\bibfnamefont {P.}~\bibnamefont {Schindler}}, \bibinfo
  {author} {\bibfnamefont {T.}~\bibnamefont {Monz}}, \bibinfo {author}
  {\bibfnamefont {P.~O.}\ \bibnamefont {Schmidt}}, \bibinfo {author}
  {\bibfnamefont {B.}~\bibnamefont {Brandst{\"a}tter}}, \bibinfo {author}
  {\bibfnamefont {T.~E.}\ \bibnamefont {Northup}}, \ and\ \bibinfo {author}
  {\bibfnamefont {R.}~\bibnamefont {Blatt}},\ }\href
  {https://doi.org/10.1038/nature11120} {\bibfield  {journal} {\bibinfo
  {journal} {Nature}\ }\textbf {\bibinfo {volume} {485}},\ \bibinfo {pages}
  {482} (\bibinfo {year} {2012})}\BibitemShut {NoStop}%
\bibitem [{\citenamefont {Stute}\ \emph {et~al.}(2013)\citenamefont {Stute},
  \citenamefont {Casabone}, \citenamefont {Brandst{\"a}tter}, \citenamefont
  {Friebe}, \citenamefont {Northup},\ and\ \citenamefont
  {Blatt}}]{Stute:2013we}%
  \BibitemOpen
  \bibfield  {author} {\bibinfo {author} {\bibfnamefont {A.}~\bibnamefont
  {Stute}}, \bibinfo {author} {\bibfnamefont {B.}~\bibnamefont {Casabone}},
  \bibinfo {author} {\bibfnamefont {B.}~\bibnamefont {Brandst{\"a}tter}},
  \bibinfo {author} {\bibfnamefont {K.}~\bibnamefont {Friebe}}, \bibinfo
  {author} {\bibfnamefont {T.~E.}\ \bibnamefont {Northup}}, \ and\ \bibinfo
  {author} {\bibfnamefont {R.}~\bibnamefont {Blatt}},\ }\href {\doibase
  10.1038/nphoton.2012.358} {\bibfield  {journal} {\bibinfo  {journal} {Nature
  Photonics}\ }\textbf {\bibinfo {volume} {7}},\ \bibinfo {pages} {219}
  (\bibinfo {year} {2013})}\BibitemShut {NoStop}%
\bibitem [{\citenamefont {Bock}\ \emph {et~al.}(2018)\citenamefont {Bock},
  \citenamefont {Eich}, \citenamefont {Kucera}, \citenamefont {Kreis},
  \citenamefont {Lenhard}, \citenamefont {Becher},\ and\ \citenamefont
  {Eschner}}]{Bock2018}%
  \BibitemOpen
  \bibfield  {author} {\bibinfo {author} {\bibfnamefont {M.}~\bibnamefont
  {Bock}}, \bibinfo {author} {\bibfnamefont {P.}~\bibnamefont {Eich}}, \bibinfo
  {author} {\bibfnamefont {S.}~\bibnamefont {Kucera}}, \bibinfo {author}
  {\bibfnamefont {M.}~\bibnamefont {Kreis}}, \bibinfo {author} {\bibfnamefont
  {A.}~\bibnamefont {Lenhard}}, \bibinfo {author} {\bibfnamefont
  {C.}~\bibnamefont {Becher}}, \ and\ \bibinfo {author} {\bibfnamefont
  {J.}~\bibnamefont {Eschner}},\ }\href
  {https://doi.org/10.1038/s41467-018-04341-2} {\bibfield  {journal} {\bibinfo
  {journal} {Nature Communications}\ }\textbf {\bibinfo {volume} {9}},\
  \bibinfo {pages} {1998} (\bibinfo {year} {2018})}\BibitemShut {NoStop}%
\bibitem [{\citenamefont {Walker}\ \emph {et~al.}(2018)\citenamefont {Walker},
  \citenamefont {Miyanishi}, \citenamefont {Ikuta}, \citenamefont {Takahashi},
  \citenamefont {Vartabi~Kashanian}, \citenamefont {Tsujimoto}, \citenamefont
  {Hayasaka}, \citenamefont {Yamamoto}, \citenamefont {Imoto},\ and\
  \citenamefont {Keller}}]{Walker2018}%
  \BibitemOpen
  \bibfield  {author} {\bibinfo {author} {\bibfnamefont {T.}~\bibnamefont
  {Walker}}, \bibinfo {author} {\bibfnamefont {K.}~\bibnamefont {Miyanishi}},
  \bibinfo {author} {\bibfnamefont {R.}~\bibnamefont {Ikuta}}, \bibinfo
  {author} {\bibfnamefont {H.}~\bibnamefont {Takahashi}}, \bibinfo {author}
  {\bibfnamefont {S.}~\bibnamefont {Vartabi~Kashanian}}, \bibinfo {author}
  {\bibfnamefont {Y.}~\bibnamefont {Tsujimoto}}, \bibinfo {author}
  {\bibfnamefont {K.}~\bibnamefont {Hayasaka}}, \bibinfo {author}
  {\bibfnamefont {T.}~\bibnamefont {Yamamoto}}, \bibinfo {author}
  {\bibfnamefont {N.}~\bibnamefont {Imoto}}, \ and\ \bibinfo {author}
  {\bibfnamefont {M.}~\bibnamefont {Keller}},\ }\href {\doibase
  10.1103/PhysRevLett.120.203601} {\bibfield  {journal} {\bibinfo  {journal}
  {Phys. Rev. Lett.}\ }\textbf {\bibinfo {volume} {120}},\ \bibinfo {pages}
  {203601} (\bibinfo {year} {2018})}\BibitemShut {NoStop}%
\bibitem [{\citenamefont {Krutyanskiy}\ \emph {et~al.}(2019)\citenamefont
  {Krutyanskiy}, \citenamefont {Meraner}, \citenamefont {Schupp}, \citenamefont
  {Krcmarsky}, \citenamefont {Hainzer},\ and\ \citenamefont {Lanyon}}]{50km}%
  \BibitemOpen
  \bibfield  {author} {\bibinfo {author} {\bibfnamefont {V.}~\bibnamefont
  {Krutyanskiy}}, \bibinfo {author} {\bibfnamefont {M.}~\bibnamefont
  {Meraner}}, \bibinfo {author} {\bibfnamefont {J.}~\bibnamefont {Schupp}},
  \bibinfo {author} {\bibfnamefont {V.}~\bibnamefont {Krcmarsky}}, \bibinfo
  {author} {\bibfnamefont {H.}~\bibnamefont {Hainzer}}, \ and\ \bibinfo
  {author} {\bibfnamefont {B.~P.}\ \bibnamefont {Lanyon}},\ }\href {\doibase
  10.1038/s41534-019-0186-3} {\bibfield  {journal} {\bibinfo  {journal} {npj
  Quantum Information}\ }\textbf {\bibinfo {volume} {5}},\ \bibinfo {pages}
  {72} (\bibinfo {year} {2019})}\BibitemShut {NoStop}%
\bibitem [{\citenamefont {Hong}\ \emph {et~al.}(1987)\citenamefont {Hong},
  \citenamefont {Ou},\ and\ \citenamefont {Mandel}}]{PhysRevLett.59.2044}%
  \BibitemOpen
  \bibfield  {author} {\bibinfo {author} {\bibfnamefont {C.~K.}\ \bibnamefont
  {Hong}}, \bibinfo {author} {\bibfnamefont {Z.~Y.}\ \bibnamefont {Ou}}, \ and\
  \bibinfo {author} {\bibfnamefont {L.}~\bibnamefont {Mandel}},\ }\href
  {\doibase 10.1103/PhysRevLett.59.2044} {\bibfield  {journal} {\bibinfo
  {journal} {Phys. Rev. Lett.}\ }\textbf {\bibinfo {volume} {59}},\ \bibinfo
  {pages} {2044} (\bibinfo {year} {1987})}\BibitemShut {NoStop}%
\bibitem [{\citenamefont {Fischer}\ \emph {et~al.}(2016)\citenamefont
  {Fischer}, \citenamefont {M{\"u}ller}, \citenamefont {Lagoudakis},\ and\
  \citenamefont {Vu{\v{c}}kovi{\'{c}}}}]{Fischer_2016}%
  \BibitemOpen
  \bibfield  {author} {\bibinfo {author} {\bibfnamefont {K.~A.}\ \bibnamefont
  {Fischer}}, \bibinfo {author} {\bibfnamefont {K.}~\bibnamefont {M{\"u}ller}},
  \bibinfo {author} {\bibfnamefont {K.~G.}\ \bibnamefont {Lagoudakis}}, \ and\
  \bibinfo {author} {\bibfnamefont {J.}~\bibnamefont {Vu{\v{c}}kovi{\'{c}}}},\
  }\href {\doibase 10.1088/1367-2630/18/11/113053} {\bibfield  {journal}
  {\bibinfo  {journal} {New Journal of Physics}\ }\textbf {\bibinfo {volume}
  {18}},\ \bibinfo {pages} {113053} (\bibinfo {year} {2016})}\BibitemShut
  {NoStop}%
\bibitem [{\citenamefont {M\"uller}\ \emph {et~al.}(2017)\citenamefont
  {M\"uller}, \citenamefont {Tentrup}, \citenamefont {Bienert}, \citenamefont
  {Morigi},\ and\ \citenamefont {Eschner}}]{PhysRevA.96.023861}%
  \BibitemOpen
  \bibfield  {author} {\bibinfo {author} {\bibfnamefont {P.}~\bibnamefont
  {M\"uller}}, \bibinfo {author} {\bibfnamefont {T.}~\bibnamefont {Tentrup}},
  \bibinfo {author} {\bibfnamefont {M.}~\bibnamefont {Bienert}}, \bibinfo
  {author} {\bibfnamefont {G.}~\bibnamefont {Morigi}}, \ and\ \bibinfo {author}
  {\bibfnamefont {J.}~\bibnamefont {Eschner}},\ }\href {\doibase
  10.1103/PhysRevA.96.023861} {\bibfield  {journal} {\bibinfo  {journal} {Phys.
  Rev. A}\ }\textbf {\bibinfo {volume} {96}},\ \bibinfo {pages} {023861}
  (\bibinfo {year} {2017})}\BibitemShut {NoStop}%
\bibitem [{\citenamefont {Legero}\ \emph {et~al.}(2004)\citenamefont {Legero},
  \citenamefont {Wilk}, \citenamefont {Hennrich}, \citenamefont {Rempe},\ and\
  \citenamefont {Kuhn}}]{PhysRevLett.93.070503}%
  \BibitemOpen
  \bibfield  {author} {\bibinfo {author} {\bibfnamefont {T.}~\bibnamefont
  {Legero}}, \bibinfo {author} {\bibfnamefont {T.}~\bibnamefont {Wilk}},
  \bibinfo {author} {\bibfnamefont {M.}~\bibnamefont {Hennrich}}, \bibinfo
  {author} {\bibfnamefont {G.}~\bibnamefont {Rempe}}, \ and\ \bibinfo {author}
  {\bibfnamefont {A.}~\bibnamefont {Kuhn}},\ }\href {\doibase
  10.1103/PhysRevLett.93.070503} {\bibfield  {journal} {\bibinfo  {journal}
  {Phys. Rev. Lett.}\ }\textbf {\bibinfo {volume} {93}},\ \bibinfo {pages}
  {070503} (\bibinfo {year} {2004})}\BibitemShut {NoStop}%
\bibitem [{\citenamefont {Wilk}\ \emph {et~al.}(2007)\citenamefont {Wilk},
  \citenamefont {Webster}, \citenamefont {Specht}, \citenamefont {Rempe},\ and\
  \citenamefont {Kuhn}}]{PhysRevLett.98.063601}%
  \BibitemOpen
  \bibfield  {author} {\bibinfo {author} {\bibfnamefont {T.}~\bibnamefont
  {Wilk}}, \bibinfo {author} {\bibfnamefont {S.~C.}\ \bibnamefont {Webster}},
  \bibinfo {author} {\bibfnamefont {H.~P.}\ \bibnamefont {Specht}}, \bibinfo
  {author} {\bibfnamefont {G.}~\bibnamefont {Rempe}}, \ and\ \bibinfo {author}
  {\bibfnamefont {A.}~\bibnamefont {Kuhn}},\ }\href {\doibase
  10.1103/PhysRevLett.98.063601} {\bibfield  {journal} {\bibinfo  {journal}
  {Phys. Rev. Lett.}\ }\textbf {\bibinfo {volume} {98}},\ \bibinfo {pages}
  {063601} (\bibinfo {year} {2007})}\BibitemShut {NoStop}%
\bibitem [{\citenamefont {Pelc}\ \emph {et~al.}(2011)\citenamefont {Pelc},
  \citenamefont {Ma}, \citenamefont {Phillips}, \citenamefont {Zhang},
  \citenamefont {Langrock}, \citenamefont {Slattery}, \citenamefont {Tang},\
  and\ \citenamefont {Fejer}}]{Pelc:11}%
  \BibitemOpen
  \bibfield  {author} {\bibinfo {author} {\bibfnamefont {J.~S.}\ \bibnamefont
  {Pelc}}, \bibinfo {author} {\bibfnamefont {L.}~\bibnamefont {Ma}}, \bibinfo
  {author} {\bibfnamefont {C.~R.}\ \bibnamefont {Phillips}}, \bibinfo {author}
  {\bibfnamefont {Q.}~\bibnamefont {Zhang}}, \bibinfo {author} {\bibfnamefont
  {C.}~\bibnamefont {Langrock}}, \bibinfo {author} {\bibfnamefont
  {O.}~\bibnamefont {Slattery}}, \bibinfo {author} {\bibfnamefont
  {X.}~\bibnamefont {Tang}}, \ and\ \bibinfo {author} {\bibfnamefont {M.~M.}\
  \bibnamefont {Fejer}},\ }\href {\doibase 10.1364/OE.19.021445} {\bibfield
  {journal} {\bibinfo  {journal} {Opt. Express}\ }\textbf {\bibinfo {volume}
  {19}},\ \bibinfo {pages} {21445} (\bibinfo {year} {2011})}\BibitemShut
  {NoStop}%
\bibitem [{Sup()}]{SuppMat}%
  \BibitemOpen
  \href@noop {} {\enquote {\bibinfo {title} {Supplementary material of this
  paper},}\ }\BibitemShut {NoStop}%
\bibitem [{\citenamefont {Keller}\ \emph {et~al.}(2004)\citenamefont {Keller},
  \citenamefont {Lange}, \citenamefont {Hayasaka}, \citenamefont {Lange},\ and\
  \citenamefont {Walther}}]{Keller:2004cf}%
  \BibitemOpen
  \bibfield  {author} {\bibinfo {author} {\bibfnamefont {M.}~\bibnamefont
  {Keller}}, \bibinfo {author} {\bibfnamefont {B.}~\bibnamefont {Lange}},
  \bibinfo {author} {\bibfnamefont {K.}~\bibnamefont {Hayasaka}}, \bibinfo
  {author} {\bibfnamefont {W.}~\bibnamefont {Lange}}, \ and\ \bibinfo {author}
  {\bibfnamefont {H.}~\bibnamefont {Walther}},\ }\href {\doibase
  10.1038/nature02961} {\bibfield  {journal} {\bibinfo  {journal} {Nature}\
  }\textbf {\bibinfo {volume} {431}},\ \bibinfo {pages} {1075} (\bibinfo {year}
  {2004})}\BibitemShut {NoStop}%
\bibitem [{\citenamefont {Krutyanskiy}\ \emph {et~al.}(2017)\citenamefont
  {Krutyanskiy}, \citenamefont {Meraner}, \citenamefont {Schupp},\ and\
  \citenamefont {Lanyon}}]{Krutyanskiy2017}%
  \BibitemOpen
  \bibfield  {author} {\bibinfo {author} {\bibfnamefont {V.}~\bibnamefont
  {Krutyanskiy}}, \bibinfo {author} {\bibfnamefont {M.}~\bibnamefont
  {Meraner}}, \bibinfo {author} {\bibfnamefont {J.}~\bibnamefont {Schupp}}, \
  and\ \bibinfo {author} {\bibfnamefont {B.~P.}\ \bibnamefont {Lanyon}},\
  }\href {\doibase 10.1007/s00340-017-6806-8} {\bibfield  {journal} {\bibinfo
  {journal} {Applied Physics B}\ }\textbf {\bibinfo {volume} {123}},\ \bibinfo
  {pages} {228} (\bibinfo {year} {2017})}\BibitemShut {NoStop}%
\bibitem [{\citenamefont {Craddock}\ \emph {et~al.}(2019)\citenamefont
  {Craddock}, \citenamefont {Hannegan}, \citenamefont {Ornelas-Huerta},
  \citenamefont {Siverns}, \citenamefont {Hachtel}, \citenamefont
  {Goldschmidt}, \citenamefont {Porto}, \citenamefont {Quraishi},\ and\
  \citenamefont {Rolston}}]{PhysRevLett.123.213601}%
  \BibitemOpen
  \bibfield  {author} {\bibinfo {author} {\bibfnamefont {A.~N.}\ \bibnamefont
  {Craddock}}, \bibinfo {author} {\bibfnamefont {J.}~\bibnamefont {Hannegan}},
  \bibinfo {author} {\bibfnamefont {D.~P.}\ \bibnamefont {Ornelas-Huerta}},
  \bibinfo {author} {\bibfnamefont {J.~D.}\ \bibnamefont {Siverns}}, \bibinfo
  {author} {\bibfnamefont {A.~J.}\ \bibnamefont {Hachtel}}, \bibinfo {author}
  {\bibfnamefont {E.~A.}\ \bibnamefont {Goldschmidt}}, \bibinfo {author}
  {\bibfnamefont {J.~V.}\ \bibnamefont {Porto}}, \bibinfo {author}
  {\bibfnamefont {Q.}~\bibnamefont {Quraishi}}, \ and\ \bibinfo {author}
  {\bibfnamefont {S.~L.}\ \bibnamefont {Rolston}},\ }\href {\doibase
  10.1103/PhysRevLett.123.213601} {\bibfield  {journal} {\bibinfo  {journal}
  {Phys. Rev. Lett.}\ }\textbf {\bibinfo {volume} {123}},\ \bibinfo {pages}
  {213601} (\bibinfo {year} {2019})}\BibitemShut {NoStop}%
\bibitem [{\citenamefont {Lamata}\ \emph {et~al.}(2011)\citenamefont {Lamata},
  \citenamefont {Leibrandt}, \citenamefont {Chuang}, \citenamefont {Cirac},
  \citenamefont {Lukin}, \citenamefont {Vuleti\ifmmode~\acute{c}\else
  \'{c}\fi{}},\ and\ \citenamefont {Yelin}}]{PhysRevLett.107.030501}%
  \BibitemOpen
  \bibfield  {author} {\bibinfo {author} {\bibfnamefont {L.}~\bibnamefont
  {Lamata}}, \bibinfo {author} {\bibfnamefont {D.~R.}\ \bibnamefont
  {Leibrandt}}, \bibinfo {author} {\bibfnamefont {I.~L.}\ \bibnamefont
  {Chuang}}, \bibinfo {author} {\bibfnamefont {J.~I.}\ \bibnamefont {Cirac}},
  \bibinfo {author} {\bibfnamefont {M.~D.}\ \bibnamefont {Lukin}}, \bibinfo
  {author} {\bibfnamefont {V.}~\bibnamefont {Vuleti\ifmmode~\acute{c}\else
  \'{c}\fi{}}}, \ and\ \bibinfo {author} {\bibfnamefont {S.~F.}\ \bibnamefont
  {Yelin}},\ }\href {\doibase 10.1103/PhysRevLett.107.030501} {\bibfield
  {journal} {\bibinfo  {journal} {Phys. Rev. Lett.}\ }\textbf {\bibinfo
  {volume} {107}},\ \bibinfo {pages} {030501} (\bibinfo {year}
  {2011})}\BibitemShut {NoStop}%
\bibitem [{\citenamefont {Casabone}\ \emph {et~al.}(2015)\citenamefont
  {Casabone}, \citenamefont {Friebe}, \citenamefont {Brandst\"atter},
  \citenamefont {Sch\"uppert}, \citenamefont {Blatt},\ and\ \citenamefont
  {Northup}}]{PhysRevLett.114.023602}%
  \BibitemOpen
  \bibfield  {author} {\bibinfo {author} {\bibfnamefont {B.}~\bibnamefont
  {Casabone}}, \bibinfo {author} {\bibfnamefont {K.}~\bibnamefont {Friebe}},
  \bibinfo {author} {\bibfnamefont {B.}~\bibnamefont {Brandst\"atter}},
  \bibinfo {author} {\bibfnamefont {K.}~\bibnamefont {Sch\"uppert}}, \bibinfo
  {author} {\bibfnamefont {R.}~\bibnamefont {Blatt}}, \ and\ \bibinfo {author}
  {\bibfnamefont {T.~E.}\ \bibnamefont {Northup}},\ }\href {\doibase
  10.1103/PhysRevLett.114.023602} {\bibfield  {journal} {\bibinfo  {journal}
  {Phys. Rev. Lett.}\ }\textbf {\bibinfo {volume} {114}},\ \bibinfo {pages}
  {023602} (\bibinfo {year} {2015})}\BibitemShut {NoStop}%
\bibitem [{\citenamefont {Steiner}\ \emph {et~al.}(2013)\citenamefont
  {Steiner}, \citenamefont {Meyer}, \citenamefont {Deutsch}, \citenamefont
  {Reichel},\ and\ \citenamefont {K\"ohl}}]{PhysRevLett.110.043003}%
  \BibitemOpen
  \bibfield  {author} {\bibinfo {author} {\bibfnamefont {M.}~\bibnamefont
  {Steiner}}, \bibinfo {author} {\bibfnamefont {H.~M.}\ \bibnamefont {Meyer}},
  \bibinfo {author} {\bibfnamefont {C.}~\bibnamefont {Deutsch}}, \bibinfo
  {author} {\bibfnamefont {J.}~\bibnamefont {Reichel}}, \ and\ \bibinfo
  {author} {\bibfnamefont {M.}~\bibnamefont {K\"ohl}},\ }\href {\doibase
  10.1103/PhysRevLett.110.043003} {\bibfield  {journal} {\bibinfo  {journal}
  {Phys. Rev. Lett.}\ }\textbf {\bibinfo {volume} {110}},\ \bibinfo {pages}
  {043003} (\bibinfo {year} {2013})}\BibitemShut {NoStop}%
\bibitem [{\citenamefont {Walker}\ \emph {et~al.}(2019)\citenamefont {Walker},
  \citenamefont {Kashanian}, \citenamefont {Ward},\ and\ \citenamefont
  {Keller}}]{walker2019improving}%
  \BibitemOpen
  \bibfield  {author} {\bibinfo {author} {\bibfnamefont {T.}~\bibnamefont
  {Walker}}, \bibinfo {author} {\bibfnamefont {S.~V.}\ \bibnamefont
  {Kashanian}}, \bibinfo {author} {\bibfnamefont {T.}~\bibnamefont {Ward}}, \
  and\ \bibinfo {author} {\bibfnamefont {M.}~\bibnamefont {Keller}},\
  }\href@noop {} {\enquote {\bibinfo {title} {Improving the
  indistinguishability of single photons from an ion-cavity system},}\ }
  (\bibinfo {year} {2019}),\ \Eprint {http://arxiv.org/abs/1911.08442}
  {arXiv:1911.08442 [quant-ph]} \BibitemShut {NoStop}%
\bibitem [{\citenamefont {{Kaiser}}\ \emph {et~al.}(2015)\citenamefont
  {{Kaiser}}, \citenamefont {{Issautier}}, \citenamefont {{Ngah}},
  \citenamefont {{Aktas}}, \citenamefont {{Delord}},\ and\ \citenamefont
  {{Tanzilli}}}]{Kaiser15}%
  \BibitemOpen
  \bibfield  {author} {\bibinfo {author} {\bibfnamefont {F.}~\bibnamefont
  {{Kaiser}}}, \bibinfo {author} {\bibfnamefont {A.}~\bibnamefont
  {{Issautier}}}, \bibinfo {author} {\bibfnamefont {L.~A.}\ \bibnamefont
  {{Ngah}}}, \bibinfo {author} {\bibfnamefont {D.}~\bibnamefont {{Aktas}}},
  \bibinfo {author} {\bibfnamefont {T.}~\bibnamefont {{Delord}}}, \ and\
  \bibinfo {author} {\bibfnamefont {S.}~\bibnamefont {{Tanzilli}}},\ }\href
  {\doibase 10.1109/JSTQE.2014.2381465} {\bibfield  {journal} {\bibinfo
  {journal} {IEEE Journal of Selected Topics in Quantum Electronics}\ }\textbf
  {\bibinfo {volume} {21}},\ \bibinfo {pages} {69} (\bibinfo {year}
  {2015})}\BibitemShut {NoStop}%
\bibitem [{\citenamefont {Maring}\ \emph {et~al.}(2017)\citenamefont {Maring},
  \citenamefont {Farrera}, \citenamefont {Kutluer}, \citenamefont {Mazzera},
  \citenamefont {Heinze},\ and\ \citenamefont {de~Riedmatten}}]{Maring2017}%
  \BibitemOpen
  \bibfield  {author} {\bibinfo {author} {\bibfnamefont {N.}~\bibnamefont
  {Maring}}, \bibinfo {author} {\bibfnamefont {P.}~\bibnamefont {Farrera}},
  \bibinfo {author} {\bibfnamefont {K.}~\bibnamefont {Kutluer}}, \bibinfo
  {author} {\bibfnamefont {M.}~\bibnamefont {Mazzera}}, \bibinfo {author}
  {\bibfnamefont {G.}~\bibnamefont {Heinze}}, \ and\ \bibinfo {author}
  {\bibfnamefont {H.}~\bibnamefont {de~Riedmatten}},\ }\href
  {https://doi.org/10.1038/nature24468} {\bibfield  {journal} {\bibinfo
  {journal} {Nature}\ }\textbf {\bibinfo {volume} {551}},\ \bibinfo {pages}
  {485} (\bibinfo {year} {2017})}\BibitemShut {NoStop}%
\bibitem [{\citenamefont {Kaiser}\ \emph {et~al.}(2019)\citenamefont {Kaiser},
  \citenamefont {Vergyris}, \citenamefont {Martin}, \citenamefont {Aktas},
  \citenamefont {Micheli}, \citenamefont {Alibart},\ and\ \citenamefont
  {Tanzilli}}]{Kaiser:19}%
  \BibitemOpen
  \bibfield  {author} {\bibinfo {author} {\bibfnamefont {F.}~\bibnamefont
  {Kaiser}}, \bibinfo {author} {\bibfnamefont {P.}~\bibnamefont {Vergyris}},
  \bibinfo {author} {\bibfnamefont {A.}~\bibnamefont {Martin}}, \bibinfo
  {author} {\bibfnamefont {D.}~\bibnamefont {Aktas}}, \bibinfo {author}
  {\bibfnamefont {M.~P.~D.}\ \bibnamefont {Micheli}}, \bibinfo {author}
  {\bibfnamefont {O.}~\bibnamefont {Alibart}}, \ and\ \bibinfo {author}
  {\bibfnamefont {S.}~\bibnamefont {Tanzilli}},\ }\href {\doibase
  10.1364/OE.27.025603} {\bibfield  {journal} {\bibinfo  {journal} {Opt.
  Express}\ }\textbf {\bibinfo {volume} {27}},\ \bibinfo {pages} {25603}
  (\bibinfo {year} {2019})}\BibitemShut {NoStop}%
\bibitem [{\citenamefont {Barros}\ \emph {et~al.}(2009)\citenamefont {Barros},
  \citenamefont {Stute}, \citenamefont {Northup}, \citenamefont {Russo},
  \citenamefont {Schmidt},\ and\ \citenamefont {Blatt}}]{Barros_2009}%
  \BibitemOpen
  \bibfield  {author} {\bibinfo {author} {\bibfnamefont {H.~G.}\ \bibnamefont
  {Barros}}, \bibinfo {author} {\bibfnamefont {A.}~\bibnamefont {Stute}},
  \bibinfo {author} {\bibfnamefont {T.~E.}\ \bibnamefont {Northup}}, \bibinfo
  {author} {\bibfnamefont {C.}~\bibnamefont {Russo}}, \bibinfo {author}
  {\bibfnamefont {P.~O.}\ \bibnamefont {Schmidt}}, \ and\ \bibinfo {author}
  {\bibfnamefont {R.}~\bibnamefont {Blatt}},\ }\href {\doibase
  10.1088/1367-2630/11/10/103004} {\bibfield  {journal} {\bibinfo  {journal}
  {New Journal of Physics}\ }\textbf {\bibinfo {volume} {11}},\ \bibinfo
  {pages} {103004} (\bibinfo {year} {2009})}\BibitemShut {NoStop}%
\bibitem [{\citenamefont {Matsukevich}\ \emph {et~al.}(2008)\citenamefont
  {Matsukevich}, \citenamefont {Maunz}, \citenamefont {Moehring}, \citenamefont
  {Olmschenk},\ and\ \citenamefont {Monroe}}]{Matsukevich2008}%
  \BibitemOpen
  \bibfield  {author} {\bibinfo {author} {\bibfnamefont {D.~N.}\ \bibnamefont
  {Matsukevich}}, \bibinfo {author} {\bibfnamefont {P.}~\bibnamefont {Maunz}},
  \bibinfo {author} {\bibfnamefont {D.~L.}\ \bibnamefont {Moehring}}, \bibinfo
  {author} {\bibfnamefont {S.}~\bibnamefont {Olmschenk}}, \ and\ \bibinfo
  {author} {\bibfnamefont {C.}~\bibnamefont {Monroe}},\ }\href {\doibase
  10.1103/PhysRevLett.100.150404} {\bibfield  {journal} {\bibinfo  {journal}
  {Phys. Rev. Lett.}\ }\textbf {\bibinfo {volume} {100}},\ \bibinfo {pages}
  {150404} (\bibinfo {year} {2008})}\BibitemShut {NoStop}%
\bibitem [{\citenamefont {da~Silva Baptista~Russo}(2008)}]{russothesis}%
  \BibitemOpen
  \bibfield  {author} {\bibinfo {author} {\bibfnamefont {C.~M.}\ \bibnamefont
  {da~Silva Baptista~Russo}},\ }\emph {\bibinfo {title} {Photon statistics of a
  single ion coupled to a high-finesse cavity}},\ \href@noop {} {Ph.D. thesis}
  (\bibinfo {year} {2008})\BibitemShut {NoStop}%
\bibitem [{\citenamefont {Roos}(2000)}]{roosthesis}%
  \BibitemOpen
  \bibfield  {author} {\bibinfo {author} {\bibfnamefont {C.}~\bibnamefont
  {Roos}},\ }\emph {\bibinfo {title} {Controlling the quantum state of trapped
  ions}},\ \href@noop {} {Ph.D. thesis} (\bibinfo {year} {2000})\BibitemShut
  {NoStop}%
\end{thebibliography}

\end{document}